\documentstyle[a4,12pt,psfig,epsf]{article}
\epsfverbosetrue
\newcommand{\be}{\begin{equation}}
\newcommand{\ee}{\end{equation}}
\newcommand{\bea}{\begin{eqnarray}}
\newcommand{\eea}{\end{eqnarray}}
\newcommand{\brr}{\begin{array}}
\newcommand{\err}{\end{array}}
\newcommand{\bc}{\begin{center}}
\newcommand{\ec}{\end{center}}

\newcommand{\ia}{a^{-1}}
\newcommand{\x}{X_{n,\nu}}
\newcommand{\xtwo}{X_{n,\nu+2}}
\def\ltap{\;\raisebox{-.5ex}{\rlap{$\sim$}} \raisebox{.5ex}{$<$}\;}
\def\gtap{\;\raisebox{-.5ex}{\rlap{$\sim$}} \raisebox{.5ex}{$>$}\;}
\newcommand{\ewxy}[2]{\setlength{\epsfxsize}{#2}\epsfbox[45 240 320 350]{#1}}
\newcommand{\B}{\multicolumn{1}{|c|}{$\bullet$}}

\begin{document}

\setcounter{page}{1}
\begin{flushright}
SWAT/129 \\
\end{flushright}

\vskip 10mm
\centerline{\Large{\bf{Lattice Monte Carlo Data versus}}}
\vskip 5mm
\centerline{\Large{\bf{Perturbation Theory.}}}

\vskip 10mm
\centerline{\bf{C.R. Allton}}

\vskip 10mm
\centerline{\em{Department of Physics, University of Wales Swansea,}}
\centerline{\em{Singleton Park, Swansea SA2 8PP, U.K.}}

\vskip 10mm
\begin{abstract}
The differences between lattice Monte Carlo data and perturbation
theory are usually associated with the ``bad'' behaviour of the
bare lattice coupling $g_0$ due to the effects of large (and unknown)
higher order coefficients in the $g_0$ perturbative series.
In this philosophy a new, renormalised coupling $g'$ is defined
with the aim of reducing the higher order coefficients of the
perturbative series in $g'$.
An improvement in the agreement between lattice data and this
new perturbation series is generally observed.

In this paper an alternative scenario is discussed where {\em lattice artifacts}
are proposed as the cause of the disagreement between lattice data
and the $g_0$-perturbative series.
We find that this interpretation provides excellent agreement
between lattice data and {\it perturbation theory in $g_0$}
corrected for lattice artifacts.
We show that this viewpoint leads typically to an order of magnitude improvement
in the agreement between lattice data and perturbation theory,
compared to typical $g'$ perturbation expansions.
The success of this procedure leads to a determination of
$\Lambda_{\overline{MS}}^{N_f=0}$ of $220 \pm 20$ MeV.
Lattice data studied includes quenched values of the string tension,
the hadronic scale $r_0$, the discrete beta function $\Delta\beta(\beta)$,
$M_\rho$, $f_\pi$ and the $1P-1S$ splitting in charmonium.
The new 3-loop term of the lattice $\beta$-function \cite{b_2} has been
incorporated in this study.

A discussion of the implication of this result for lattice calculations
is presented.
\end{abstract}

\newpage

\section{Introduction}
\label{sec:intro}

A necessary condition for lattice predictions of QCD and other
asymptotically free theories to have physical (continuum) relevance is
that they reproduce weak coupling perturbation theory (PT) in the limit
of the bare coupling $g_0\rightarrow 0$. This perturbative scaling
behaviour (a.k.a. asymptotic scaling) has not yet been observed for
complicated theories like QCD. Even for simple systems such as the
2-dimensional $O(3)$ model, there is evidence that in present simulations,
the correlation length, $\xi$, is still 4\% away from the perturbative
behaviour even for huge values of $\xi \sim {\cal O}(10^5)$
\cite{sokal}.

These comments apply when the ``naive'', bare lattice coupling
from the lattice action, $g_0$, is used as the expansion parameter
in the perturbative series.
As a result of this disappointing disagreement, various workers
have proposed methods of improving the convergence of the
perturbation series by re-expanding it in terms of some new coupling $g'$
\cite{parisi,lm}.
They argue that the lack of (2-loop) perturbative scaling
is due to the presence of higher order terms
in the perturbation expansion appearing with large coefficients.
If a more physical coupling, $g'$, could be found, then, it is argued that
the perturbation series expressed in $g'$ will converge faster
(i.e. it will have smaller higher order terms).
There has been evidence to support this philosophy
(see for example \cite{lm,wup1}).
We refer to this method as the ``renormalised coupling'' approach.

This paper studies an alternative viewpoint in which the above disagreement
stems from lattice artifacts, i.e. cut-off effects \cite{cra}.
All lattice Monte Carlo data is
obtained on a lattice with finite lattice spacing, $a$,
with an action that is correct to ${\cal O}(a^n)$ for some $n$.
Therefore, {\it a priori}, corrections of the form of ${\cal O}(a^n)$ at best,
are present in any observable.
These terms are usually treated in the final stage of the analysis,
when an extrapolation to the continuum limit of some {\em physical}
quantity is performed (see for example \cite{wup1,aida,lat93_claude}).
In this paper, these terms are shown to provide the mismatch
between lattice Monte Carlo data, expressed in lattice units, and PT
without resorting to the use of a renormalised coupling $g'$.
That is, bare lattice data is reproduced by $g_0$-PT
with the simple addition of terms of ${\cal O}(a^n)$, with an appropriate
value for $n$.
We call this agreement between lattice data and PT
``Lattice-Distorted Perturbative Scaling''.

The QCD quantities studied in this analysis are:
the string tension, $\sqrt\sigma$;
the hadronic scale, $r_0$ \cite{sommer};
$M_\rho$; $f_\pi$; the $1P-1S$ splitting in charmonium \cite{aida};
and the discrete beta function $\Delta\beta$.
We find that the lattice-distorted PT
reproduces the Monte Carlo data for all the quantities considered.
When fits are performed to $\sqrt\sigma$, $r_0$ and $\Delta\beta$
using various proposed renormalised coupling
schemes, the $\chi^2/dof$ are around an order of magnitude worse
than those obtained with lattice-distorted PT.
\footnote{In the case of $M_\rho$, $f_\pi$ and the $1P-1S$ splitting,
the lattice data is too noisy to constrain the fits.}
The implication of this result is that the higher order terms
in the $g_0$-PT expansion of these quantities {\it do not}
``contaminate'' the data, whereas the ${\cal O}(a)$-type terms {\it do}.

As this paper was in the final stages of preparation, a calculation
of the 3-loop coefficient of the lattice $\beta$-function appeared
\cite{b_2}. We has included this term in this study where appropriate.

The plan of this paper is as follows.
In the next section, the
(straightforward) formalism of lattice-distorted PT is presented.
Section \ref{sec:ldpt_fit} fits the QCD lattice data to
lattice-distorted PT and
obtains a value of $\Lambda_{\overline{MS}}^{N_f=0} = 220 \pm 20$ MeV.
Section \ref{sec:g'} fits the same data to the renormalised coupling-PT
schemes, specifically the $g_{\overline{MS}}$
scheme of \cite{aida}, the $g_V$-PT scheme of \cite{lm}, and the
$g_E$-PT schemes of \cite{parisi} \& \cite{wup1}.
We conclude with an interpretation of these
results, and comment on how this method can be used to
find the continuum value of quantities determined on the lattice.

A brief account of the ideas in this paper appears in \cite{cra_lat96}.

\section{Lattice-Distorted Perturbation Theory}
\label{sec:ldpt_theory}

We begin this discussion with one of the fundamental
quantities in perturbative field theory, the (3-loop) beta function:
\[
\beta(g^2) = -a \frac{dg^2}{da} = -2 b_0 g^4 - 2 b_1 g^6 - 2 b_2 g^8,
\]
where the one- and two-loop coefficients for the quenched theory are
\[
b_0 = \frac {11} {(4\pi)^2},  \,\,\,\,\,\,\,\,\,\,\,
b_1 = \frac {102}{(4\pi)^4}.
\]
The (scheme-dependent) three-loop coefficient has recently been
calculated \cite{b_2} for the $SU(N)$ lattice action,
\[
b^L_2 = \left( \frac{N}{16\pi^2} \right)^3
\left( -366.2 + \frac{1433.8}{N^2} - \frac{2143.}{N^4} \right).
\]
The $\beta$-function can be intergrated to give the usual
form for the running of the coupling with the U.V.
cut-off $1/a$,
\be
\ia = \frac{\Lambda}{f_{PT}(g^2)},
\label{eq:ia_PT}
\ee
where\footnote{
Note that the definition of $f_{PT}$ differs from that in \cite{cra} and
\cite{cra_lat96} by a factor of $b_0^{-b_1\over 2b_0^2}$ to conform with
convention.}

\be
f_{PT}(g^2) = e^{- \frac{1}{2 b_0 g^2} } \;\; (b_0 g^2)^{-b_1 \over {2 b_0^2}}
 \;\; ( 1 + d_2 g^2 ),
\label{eq:f_PT}
\ee
where $d_2 = \frac{1}{2b_0^3}(b_1^2 - b_2 b_0)$. In the case of the
lattice scheme, we have $d^L_2 = \frac{1}{2b_0^3}(b_1^2 - b^L_2 b_0) =
0.1896$ \cite{b_2}.

\begin{figure}

\setlength{\unitlength}{0.240900pt}
\ifx\plotpoint\undefined\newsavebox{\plotpoint}\fi
\sbox{\plotpoint}{\rule[-0.200pt]{0.400pt}{0.400pt}}%
\special{em:linewidth 0.4pt}%
\begin{picture}(1500,900)(0,0)
\font\gnuplot=cmr10 at 10pt
\gnuplot
\put(220,113){\special{em:moveto}}
\put(240,113){\special{em:lineto}}
\put(1436,113){\special{em:moveto}}
\put(1416,113){\special{em:lineto}}
\put(198,113){\makebox(0,0)[r]{4}}
\put(220,240){\special{em:moveto}}
\put(240,240){\special{em:lineto}}
\put(1436,240){\special{em:moveto}}
\put(1416,240){\special{em:lineto}}
\put(198,240){\makebox(0,0)[r]{4.5}}
\put(220,368){\special{em:moveto}}
\put(240,368){\special{em:lineto}}
\put(1436,368){\special{em:moveto}}
\put(1416,368){\special{em:lineto}}
\put(198,368){\makebox(0,0)[r]{5}}
\put(220,495){\special{em:moveto}}
\put(240,495){\special{em:lineto}}
\put(1436,495){\special{em:moveto}}
\put(1416,495){\special{em:lineto}}
\put(198,495){\makebox(0,0)[r]{5.5}}
\put(220,622){\special{em:moveto}}
\put(240,622){\special{em:lineto}}
\put(1436,622){\special{em:moveto}}
\put(1416,622){\special{em:lineto}}
\put(198,622){\makebox(0,0)[r]{6}}
\put(220,750){\special{em:moveto}}
\put(240,750){\special{em:lineto}}
\put(1436,750){\special{em:moveto}}
\put(1416,750){\special{em:lineto}}
\put(198,750){\makebox(0,0)[r]{6.5}}
\put(220,877){\special{em:moveto}}
\put(240,877){\special{em:lineto}}
\put(1436,877){\special{em:moveto}}
\put(1416,877){\special{em:lineto}}
\put(198,877){\makebox(0,0)[r]{7}}
\put(292,113){\special{em:moveto}}
\put(292,133){\special{em:lineto}}
\put(292,877){\special{em:moveto}}
\put(292,857){\special{em:lineto}}
\put(292,68){\makebox(0,0){5.4}}
\put(435,113){\special{em:moveto}}
\put(435,133){\special{em:lineto}}
\put(435,877){\special{em:moveto}}
\put(435,857){\special{em:lineto}}
\put(435,68){\makebox(0,0){5.6}}
\put(578,113){\special{em:moveto}}
\put(578,133){\special{em:lineto}}
\put(578,877){\special{em:moveto}}
\put(578,857){\special{em:lineto}}
\put(578,68){\makebox(0,0){5.8}}
\put(721,113){\special{em:moveto}}
\put(721,133){\special{em:lineto}}
\put(721,877){\special{em:moveto}}
\put(721,857){\special{em:lineto}}
\put(721,68){\makebox(0,0){6}}
\put(864,113){\special{em:moveto}}
\put(864,133){\special{em:lineto}}
\put(864,877){\special{em:moveto}}
\put(864,857){\special{em:lineto}}
\put(864,68){\makebox(0,0){6.2}}
\put(1007,113){\special{em:moveto}}
\put(1007,133){\special{em:lineto}}
\put(1007,877){\special{em:moveto}}
\put(1007,857){\special{em:lineto}}
\put(1007,68){\makebox(0,0){6.4}}
\put(1150,113){\special{em:moveto}}
\put(1150,133){\special{em:lineto}}
\put(1150,877){\special{em:moveto}}
\put(1150,857){\special{em:lineto}}
\put(1150,68){\makebox(0,0){6.6}}
\put(1293,113){\special{em:moveto}}
\put(1293,133){\special{em:lineto}}
\put(1293,877){\special{em:moveto}}
\put(1293,857){\special{em:lineto}}
\put(1293,68){\makebox(0,0){6.8}}
\put(1436,113){\special{em:moveto}}
\put(1436,133){\special{em:lineto}}
\put(1436,877){\special{em:moveto}}
\put(1436,857){\special{em:lineto}}
\put(1436,68){\makebox(0,0){7}}
\put(220,113){\special{em:moveto}}
\put(1436,113){\special{em:lineto}}
\put(1436,877){\special{em:lineto}}
\put(220,877){\special{em:lineto}}
\put(220,113){\special{em:lineto}}
\put(45,495){\makebox(0,0){$\Lambda_\sigma$ [MeV]}}
\put(828,23){\makebox(0,0){$\beta$}}
\put(506,172){\raisebox{-.8pt}{\makebox(0,0){$\Diamond$}}}
\put(578,287){\raisebox{-.8pt}{\makebox(0,0){$\Diamond$}}}
\put(649,393){\raisebox{-.8pt}{\makebox(0,0){$\Diamond$}}}
\put(721,475){\raisebox{-.8pt}{\makebox(0,0){$\Diamond$}}}
\put(864,629){\raisebox{-.8pt}{\makebox(0,0){$\Diamond$}}}
\put(1007,718){\raisebox{-.8pt}{\makebox(0,0){$\Diamond$}}}
\put(1293,796){\raisebox{-.8pt}{\makebox(0,0){$\Diamond$}}}
\put(1078,740){\raisebox{-.8pt}{\makebox(0,0){$\Diamond$}}}
\put(721,546){\raisebox{-.8pt}{\makebox(0,0){$\Diamond$}}}
\put(864,637){\raisebox{-.8pt}{\makebox(0,0){$\Diamond$}}}
\put(1007,715){\raisebox{-.8pt}{\makebox(0,0){$\Diamond$}}}
\put(506,169){\special{em:moveto}}
\put(506,176){\special{em:lineto}}
\put(496,169){\special{em:moveto}}
\put(516,169){\special{em:lineto}}
\put(496,176){\special{em:moveto}}
\put(516,176){\special{em:lineto}}
\put(578,282){\special{em:moveto}}
\put(578,293){\special{em:lineto}}
\put(568,282){\special{em:moveto}}
\put(588,282){\special{em:lineto}}
\put(568,293){\special{em:moveto}}
\put(588,293){\special{em:lineto}}
\put(649,375){\special{em:moveto}}
\put(649,411){\special{em:lineto}}
\put(639,375){\special{em:moveto}}
\put(659,375){\special{em:lineto}}
\put(639,411){\special{em:moveto}}
\put(659,411){\special{em:lineto}}
\put(721,441){\special{em:moveto}}
\put(721,509){\special{em:lineto}}
\put(711,441){\special{em:moveto}}
\put(731,441){\special{em:lineto}}
\put(711,509){\special{em:moveto}}
\put(731,509){\special{em:lineto}}
\put(864,611){\special{em:moveto}}
\put(864,646){\special{em:lineto}}
\put(854,611){\special{em:moveto}}
\put(874,611){\special{em:lineto}}
\put(854,646){\special{em:moveto}}
\put(874,646){\special{em:lineto}}
\put(1007,702){\special{em:moveto}}
\put(1007,734){\special{em:lineto}}
\put(997,702){\special{em:moveto}}
\put(1017,702){\special{em:lineto}}
\put(997,734){\special{em:moveto}}
\put(1017,734){\special{em:lineto}}
\put(1293,767){\special{em:moveto}}
\put(1293,825){\special{em:lineto}}
\put(1283,767){\special{em:moveto}}
\put(1303,767){\special{em:lineto}}
\put(1283,825){\special{em:moveto}}
\put(1303,825){\special{em:lineto}}
\put(1078,725){\special{em:moveto}}
\put(1078,754){\special{em:lineto}}
\put(1068,725){\special{em:moveto}}
\put(1088,725){\special{em:lineto}}
\put(1068,754){\special{em:moveto}}
\put(1088,754){\special{em:lineto}}
\put(721,529){\special{em:moveto}}
\put(721,563){\special{em:lineto}}
\put(711,529){\special{em:moveto}}
\put(731,529){\special{em:lineto}}
\put(711,563){\special{em:moveto}}
\put(731,563){\special{em:lineto}}
\put(864,620){\special{em:moveto}}
\put(864,655){\special{em:lineto}}
\put(854,620){\special{em:moveto}}
\put(874,620){\special{em:lineto}}
\put(854,655){\special{em:moveto}}
\put(874,655){\special{em:lineto}}
\put(1007,682){\special{em:moveto}}
\put(1007,748){\special{em:lineto}}
\put(997,682){\special{em:moveto}}
\put(1017,682){\special{em:lineto}}
\put(997,748){\special{em:moveto}}
\put(1017,748){\special{em:lineto}}
\end{picture}

\caption{\it Plot of $\Lambda_\sigma = \ia_\sigma \; f_{PT}(g_0^2)$
against $\beta = 6/g_0^2$. The 3-loop form of $f_{PT}$ was used.
\label{fig:lambda_sigma}}
\end{figure}
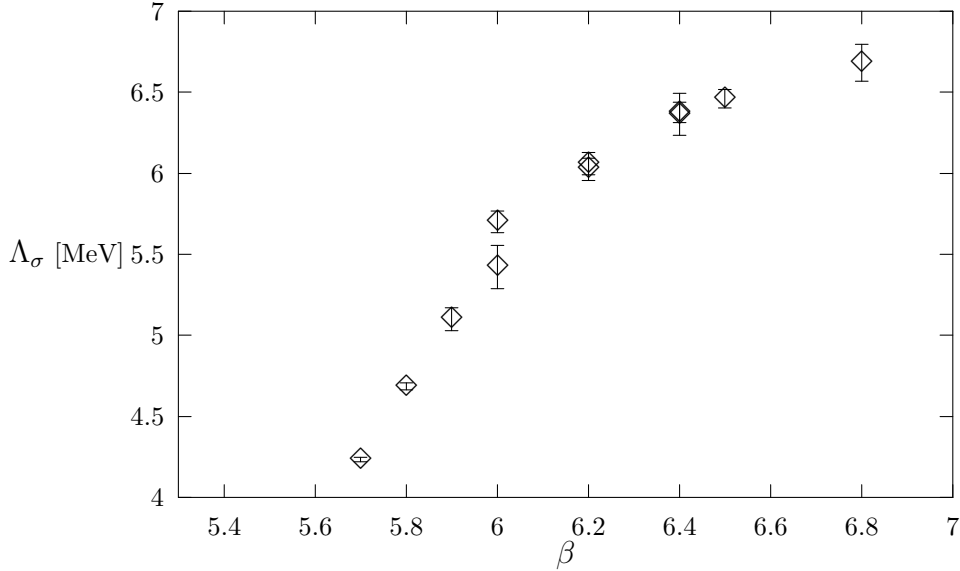

All coupling and fields on the lattice, and therefore
all observables, are dimensionless.
Lattice calculations set the scale $a^{-1}$
by calculating some quantity on the lattice,
eg. the string tension, $\sigma_L$,
and comparing it with its experimental (dimensionful) value,
$\sigma_{exp}$:
\[
\ia_{\sigma} = \frac{ \sqrt{\sigma_{exp}} } {\sqrt{\sigma_L}}
\]
Typical values for $\ia_{\sigma}$ from recent simulations are listed
in the third column of table \ref{tab:inva} together the references
and corresponding $\beta=6/g_0^2$ values in columns 1 and 2.
It is now easy to check if $\ia_\sigma$ follows $g_0$-PT
(i.e. eq.(\ref{eq:ia_PT}) with $g=g_0$).
Plotting $\Lambda_\sigma = \ia_{\sigma} f_{PT}(g_0^2)$ in fig
\ref{fig:lambda_sigma}, we observe a non-constant behaviour,
signaling the failure of Monte Carlo data to
follow even 3-loop $g_0$-PT.
(The failure of 2-loop perturbative running has already long been noted.)

There are a number of possible causes of the disagreement.

\begin{itemize}
\item quenching
\item finite volume effects
\item unphysically large value of the quark mass (relevant for
      ``hadronic'' quantities such as masses and decay constants,
      rather than for $\sigma$ as used in the above example)
\item a real non-perturbative effect
\item the inclusion of only a finite number of terms in the PT expansion
\item lattice artifacts due to the finiteness of $a$
\end{itemize}

For the reasons outlined in \cite{cra}, the first three
effects cannot give rise to the sizeable discrepancy between
lattice data and PT in fig \ref{fig:lambda_sigma}.
(For example, quenching should modify
$\Lambda$ by an overall {\it constant} factor.)
As far as true (i.e. continuum) non-perturbative effects are concerned,
the overwhelming expectation is that for cut-offs of
$\ia \gtap 2$ GeV these effects should be minimal.
(In any case these effects have the same form
as lattice artifacts since they are of the form $e^{-1/g^2} \sim {\cal O}(a)$.)
Therefore, we can assume that the disagreement is due to either or both
of the last two possibilities.

The effects of higher orders in $g_0$ and the finiteness
of $a$ can be parametrised in the lattice beta function as follows:
\bea\nonumber
\beta_L(g_0^2) &=& -a_L \frac{dg_0^2}{da_L} =
-(2 b_0 g_0^4 + 2 b_1 g_0^6 + 2 b^L_2 g_0^8 + \sum_{l=4} b^L_{l-1} g_0^{2l+2}) \\
&\times&  ( 1 + \sum_{n=1} c_n(g_0^2) a_L^n(g_0^2) )
\label{eq:beta_ldpt}
\eea
Here the $b^L_3, b^L_4, ...$ are the (unknown) higher loop coefficients of the
lattice beta function.
(They are presumed to be large in the renormalised coupling approach.)
The $c_n$ are the (non-universal) coefficients of the ${\cal O}(a^n)$
pieces and are, in general, polynomial functions of $g_0^2$.\footnote{
We have used the replacement $g_0^2 \log a \sim 1$ in
eq.(\ref{eq:beta_ldpt}) since the difference between $g_0^2$ and
$\log^{-1}a$ can be incorporated into the higher order terms.}

Eq.(\ref{eq:beta_ldpt}) can be integrated giving
\be
\ia_L(g_0^2) = \frac{\Lambda_L}{f_{PT}(g_0^2)}
    \times  ( 1 + \sum_{l=4} d^L_{l-1} g_0^{2l-4} )^{-1}
    \times  ( 1 + \sum_{n=1} c'_n(g_0^2) f_{PT}^{\;n}(g_0^2) )
\label{eq:ia_ldpt}
\ee
Note that the ${\cal O}(a^n)$ term has been expressed in terms of $f_{PT}^{\;n}$.
This can be done without any loss of generality since any difference
between $a_L$ and $f_{PT}$ is higher order and can be absorbed
into the coefficients $c'_m(g_0^2)$ for $m\ge n$.

In the following section we study lattice-distorted PT by setting the
higher order coefficients, $d^L$, to zero,
and fitting the data in order to determine the $c'_n$.
We perform this fit for both the 2-loop function (i.e. we set $d_2^L$ to
zero in $f_{PT}$) and the 3-loop function.
In section \ref{sec:g'} the renormalised coupling ideas are
studied: i.e. all the $c'_n$ are set to zero, and 
$g_0$ is replaced by $g_{\overline{MS}}$ \cite{aida}, $g_V$ \cite{lm},
$g_E$ \cite{parisi} and $g_{E2}$ \cite{wup1}.
In these cases, we also fit to the appropriate 2-loop and 3-loop formulae.

\begin{table}
\begin{center}
\begin{tabular}{lllllll}
\hline
\hline
Ref.                 & $\beta=6/g^2$
                            & $a^{-1}_\sigma$
                                       & $a^{-1}_{r_0}$
                                                 & $a^{-1}_{M_\rho}$
                                                            & $a^{-1}_{f_\pi}$
                                                                        & $a^{-1}_{1P-1S}$ \\
%
\hline
\cite{bali} - WUP    & 5.5  &          & 0.80(1) &          &           & \\
\cite{bali} - WUP    & 5.6  &          & 0.91(2) &          &           & \\
\cite{born} - MT$_c$ & 5.7  & 1.073(3) &         &          &           &         \\
\cite{bali} - WUP    & 5.7  &          & 1.14(2) &          &           & \\
\cite{ibm_new} - GF11& 5.7  &          &         & 1.42(2)  & 1.08(5)   & \\
\cite{aida} - FNAL   & 5.7  &          &         &          &           & 1.15(8) \\
\cite{wup2} - WPC    & 5.74 &          &         & 1.44(3)  &           & \\
\cite{born} - MT$_c$ & 5.8  & 1.333(6) &         &          &           & \\
\cite{bali} - WUP    & 5.8  &          & 1.45(2) &          &           & \\
\cite{born} - MT$_c$ & 5.9  & 1.63(2)  &         &          &           & \\
\cite{bali} - WUP    & 5.9  &          & 1.85(5) &          &           & \\
\cite{aida} - FNAL   & 5.9  &          &         &          &           & 1.78(9) \\
\cite{ibm_new} - GF11& 5.93 &          &         & 1.99(4)  & 1.78(5)   & \\
\cite{wup2} - WPC    & 6.0  &          &         & 2.25(10) &           & \\
\cite{wup1} - WUP    & 6.0  & 1.94(5)  &         &          &           & \\
\cite{ukqcd2} - UKQCD& 6.0  & 2.04(2)  &         &          &           & \\
\cite{bali} - WUP    & 6.0  &          & 2.11(2) &          &           & \\
\cite{ukqcd2} - UKQCD& 6.0  &          & 2.19(4) &          &           & \\
\cite{ape_6018} - APE& 6.0  &          &         & 2.23(5)  & 1.98(8)   & \\
\cite{ape_dallas} - APE&6.0 &          &         & 2.18(9)  & 1.76(8)   & \\
\cite{ape_biel} - APE& 6.1  &          &         & 2.64(16) & 2.57(15)  & \\
\cite{aida} - FNAL   & 6.1  &          &         &          &           & 2.43(15) \\
\cite{ibm_new} - GF11& 6.17 &          &         & 2.77(4)  & 2.56(7)   & \\
\cite{wup1} - WUP    & 6.2  & 2.72(3)  &         &          &           & \\
\cite{ukqcd2} - UKQCD& 6.2  & 2.73(3)  &         &          &           & \\
\cite{bali} - WUP    & 6.2  &          & 2.94(2) &          &           & \\
\cite{ukqcd2} - UKQCD& 6.2  &          & 2.92(6) &          &           & \\
\cite{ape_dallas} - APE&6.2 &          &         & 2.88(24) & 2.69(24)  & \\
\cite{wup2} - WPC    & 6.26 &          &         & 3.69(32) &           & \\
\cite{wup1} - WUP    & 6.4  & 3.62(4)  &         &          &           & \\
\cite{bali} - WUP    & 6.4  &          & 3.95(3) &          &           & \\
\cite{abada} - ELC   & 6.4  &          &         & 3.70(15) & 3.7(6)    & \\
\cite{ape_biel} - APE& 6.4  &          &         & 4.09(18) & 3.48(16)  & \\
\cite{ukqcd2} - UKQCD& 6.4  & 3.62(7)  &         &          &           & \\
\cite{ukqcd2} - UKQCD& 6.4  &          & 3.90(6) &          &           & \\
\cite{ukqcd1} - UKQCD& 6.5  & 4.12(4)  &         &          &           & \\
\cite{wup1} - WUP    & 6.8  & 6.0(1)   &         &          &           & \\
\cite{bali} - WUP    & 6.8  &          & 6.7(2)  &          &           & \\
\hline
\hline
\end{tabular}
\caption{ \it{Values for $a^{-1}$ obtained from various group's work
using the Wilson action.
Note that we have used $\sigma_{exp}= (440$ MeV$)^2$, and $r_0^{-1} =
400$ MeV.}
\label{tab:inva}}
\end{center}
\end{table}

\begin{table}
\begin{center}
\begin{tabular}{|l|l|l|l|l|l|l|}
\hline 
\hline 
        & \multicolumn{5}{|c|}{\mbox{\boldmath$a^{-1}$} {\bf from}} & \\
          \cline{2-6}
	& \multicolumn{1}{|c|}{\mbox{\boldmath$\sqrt\sigma$}}
                     & \multicolumn{1}{|c|}{\mbox{\boldmath$r_0$}}
                                & \multicolumn{1}{|c|}{\mbox{\boldmath$M_\rho$}}
                                                & \multicolumn{1}{|c|}{\mbox{\boldmath$f_\pi$}}
                                                          & \multicolumn{1}{|c|}{\mbox{\boldmath$1P-1S$}}
                                                                    & \multicolumn{1}{|c|}{\mbox{\boldmath$\Delta\beta(\beta)$}} \\
%
\hline 
\hline 
\multicolumn{7}{|c|}{} \\
\multicolumn{7}{|c|}{\mbox{\boldmath$g_0$}\bf-Perturbation Theory} \\
\hline
$\Lambda_L$ [MeV]& 3.856(8) & 4.91(2)   & 5.00(4) & 4.54(6) & 4.5(2) & --- \\
$\chi^2/dof$     & 484      & 262       & 10      & 9       & 6      & 702 \\
%
\hline 
\hline 
\multicolumn{7}{|c|}{} \\
\multicolumn{7}{|c|}{\bf Leading-Order Lattice Distorted PT}\\
\multicolumn{7}{|c|}{i.e. Fit using eqs.(\ref{eq:ldpt_fit},\ref{eq:alphan})}\\
\hline
$\Lambda_L$ [MeV] & 5.85(3)  & 6.02(3)   & 6.6(2)  & 6.9(3)  & 7.6(9)  & --- \\
$\x$              & 0.204(2) & 0.150(2)  & 0.22(2) & 0.34(3) & 0.35(6) & 0.195(2) \\
$\chi^2/dof$      & 3        & 16        & 1.1     & 1.6     & 0.3     & 3.5 \\
%
\hline 
\multicolumn{7}{|c|}{\bf Next-to-Leading-Order Lattice Distorted PT} \\
\multicolumn{7}{|c|}{i.e. Fit using eqs.(\ref{eq:ldpt_fit2},\ref{eq:alphan})}\\
\hline
$\Lambda_L$ [MeV]& 6.02(5)  & 6.58(5)   & ---     & ---     & ---     & ---      \\
$\x$             & 0.26(2)  & 0.29(1)   & ---     & ---     & ---     & 0.252(6) \\
$\xtwo$          &-0.024(6) &-0.046(3)  & ---     & ---     & ---     & -0.025(3) \\
$\chi^2/dof$     & 1.7      & 1.4       & ---     & ---     & ---     & 1.7      \\
%
\hline 
\hline 
\multicolumn{7}{|c|}{} \\
\multicolumn{7}{|c|}{\mbox{\boldmath$g_{\overline{MS}}$}\bf-Perturbation Theory} \\
\hline
$\Lambda_{\overline{MS}}$ [MeV]
                & 53.3(1) & 64.2(2)  & 65.7(5) & 64.5(8) & 59(2)  & --- \\
$\chi^2/dof$    & 160     & 47       & 1.3     & 2.5     & 1.5    & 78 \\
%
\hline 
\hline 
\multicolumn{7}{|c|}{} \\
\multicolumn{7}{|c|}{\mbox{\boldmath$g_V^{(I)}$}\bf-Perturbation Theory} \\
\hline
$\Lambda'_V$ [MeV]    & 81.2(2) & 98.2(3)  & 100.3(7) & 99(1)  & 91(3)  & --- \\
$\chi^2/dof$         & 176     & 54       & 1.4      & 2.7    & 1.6    & 89 \\
\hline 
\hline 
\multicolumn{7}{|c|}{} \\
\multicolumn{7}{|c|}{\mbox{\boldmath$g_V^{(II)}$}\bf-Perturbation Theory} \\
\hline
$\Lambda'_V$ [MeV]    & 104.3(2) & 118.9(4)  & 123.9(9) & 119(2)  & 113(4) & --- \\
$\chi^2/dof$         & 31       & 13        & 5.2      & 1.4     & 0.13   & 11 \\
%
\hline 
\hline 
\multicolumn{7}{|c|}{} \\
\multicolumn{7}{|c|}{\mbox{\boldmath$g_E$}\bf-Perturbation Theory} \\
\hline
$\Lambda_E$ [MeV]& 14.80(3) & 17.09(6) & 17.7(1) & 17.1(2) & 16.1(6) & --- \\
$\chi^2/dof$     & 52       & 15       & 3.6     & 1.4     & 0.3     & 19 \\
%
\hline 
\multicolumn{7}{|c|}{} \\
\multicolumn{7}{|c|}{\mbox{\boldmath$g_{E2}$}\bf-Perturbation Theory} \\
\hline
$\Lambda_E$ [MeV]& 8.07(2) & 9.25(3) & 9.62(7) & 9.3(1)  & 8.8(3)  & --- \\
$\chi^2/dof$     & 39      & 12      & 4.4     & 1.4     & 0.2     & 13 \\
%
\hline 
\hline 
\end{tabular}
\caption{ \it{Fits of lattice data to 2-loop PT using i) $g_0$-PT,
ii) Leading Order Lattice Distorted PT,
iii) Next-to-Leading Order Lattice Distorted PT,
iv) $g_{\overline MS}$-PT,
v) $g_V^{(I,II)}$-PT,
vi) $g_E$-PT and
vii) $g_{E2}$-PT.}}
\label{tab:2fit}
\end{center}
\end{table}

\begin{table}
\begin{center}
\begin{tabular}{|l|l|l|l|l|l|l|}
\hline 
\hline 
        & \multicolumn{5}{|c|}{\mbox{\boldmath$a^{-1}$} {\bf from}} & \\
          \cline{2-6}
	& \multicolumn{1}{|c|}{\mbox{\boldmath$\sqrt\sigma$}}
                     & \multicolumn{1}{|c|}{\mbox{\boldmath$r_0$}}
                                & \multicolumn{1}{|c|}{\mbox{\boldmath$M_\rho$}}
                                                & \multicolumn{1}{|c|}{\mbox{\boldmath$f_\pi$}}
                                                          & \multicolumn{1}{|c|}{\mbox{\boldmath$1P-1S$}}
                                                                    & \multicolumn{1}{|c|}{\mbox{\boldmath$\Delta\beta(\beta)$}} \\
%
\hline 
\hline 
\multicolumn{7}{|c|}{} \\
\multicolumn{7}{|c|}{\mbox{\boldmath$g_0$}\bf-Perturbation Theory} \\
\hline
$\Lambda_L$ [MeV]  & 4.62(1)  & 5.85(2)   & 5.96(4) & 5.40(8) & 5.3(2)  & ---   \\
$\chi^2/dof$       & 448      & 239       & 9       & 8       & 5       & 625 \\
%
\hline 
\hline 
\multicolumn{7}{|c|}{} \\
\multicolumn{7}{|c|}{\bf Leading-Order Lattice Distorted PT}\\
\multicolumn{7}{|c|}{i.e. Fit using eqs.(\ref{eq:ldpt_fit},\ref{eq:alphan})}\\
\hline
$\Lambda_L$ [MeV]& 6.86(4)  & 7.07(3)   & 7.7(2)  & 8.0(4)  & 8(1)    & --- \\
$\x$             & 0.193(2) & 0.141(2)  & 0.20(2) & 0.32(3) & 0.34(6) & 0.184(2) \\
$\chi^2/dof$     & 3        & 15        & 1.1     & 1.6     & 0.3     & 3 \\
%
\hline 
\multicolumn{7}{|c|}{\bf Next-to-Leading-Order Lattice Distorted PT} \\
\multicolumn{7}{|c|}{i.e. Fit using eqs.(\ref{eq:ldpt_fit2},\ref{eq:alphan})}\\
\hline
$\Lambda_L$ [MeV]& 7.01(6)  & 7.68(6)   & ---     & ---     & ---     & ---      \\
$\x$             & 0.24(1)  & 0.27(1)   & ---     & ---     & ---     & 0.23(1) \\
$\xtwo$          &-0.019(6) &-0.042(3)  & ---     & ---     & ---     & -0.021(4) \\
$\chi^2/dof$     & 1.8      & 1.4       & ---     & ---     & ---     & 1.6      \\
%
\hline 
\hline 
\multicolumn{7}{|c|}{} \\
\multicolumn{7}{|c|}{\mbox{\boldmath$g_E$}\bf-Perturbation Theory} \\
\hline
$\Lambda_E$ [MeV]& 17.02(4) & 19.48(6) & 20.3(2) & 19.6(3) & 18.5(6) & --- \\
$\chi^2/dof$     & 36       & 12       & 5       & 1.4     & 0.18    & 12 \\
%
\hline 
\hline 
\end{tabular}
\caption{ \it{Fits of lattice data using 3-loop PT to i) $g_0$-PT,
ii) Leading Order Lattice Distorted PT,
iii) Next-to-Leading Order Lattice Distorted PT, and
iv) $g_E$-PT.}
}
\label{tab:3fit}
\end{center}
\end{table}

\begin{table}
\begin{center}
\begin{tabular}{|l|l|l|l|l|l|l|}
\hline 
\hline 
        & \multicolumn{5}{|c|}{\mbox{\boldmath$a^{-1}$} {\bf from}} & \\
          \cline{2-6}
	& \multicolumn{1}{|c|}{\mbox{\boldmath$\sqrt\sigma$}}
                     & \multicolumn{1}{|c|}{\mbox{\boldmath$r_0$}}
                                & \multicolumn{1}{|c|}{\mbox{\boldmath$M_\rho$}}
                                                & \multicolumn{1}{|c|}{\mbox{\boldmath$f_\pi$}}
                                                          & \multicolumn{1}{|c|}{\mbox{\boldmath$1P-1S$}}
                                                                    & \multicolumn{1}{|c|}{\mbox{\boldmath$\Delta\beta(\beta)$}} \\
%
\hline 
\hline 
\multicolumn{7}{|c|}{} \\
\multicolumn{7}{|c|}{\mbox{\boldmath$g_{\overline{MS}}$}\bf-Perturbation Theory} \\
\hline
$\Lambda_{\overline{MS}}$ [MeV]
                & 111(1)   & 104(2)  & 69(6)   & 103(13) & 120(30) &
                                           \multicolumn{1}{|c|}{---} \\
$d^{\overline{MS}}_2$
                & 0.483(6) & 0.39(1) & 0.04(8) & 0.37(8) & 0.47(14)&  \B       \\
$\chi^2/dof$    & 4        & 10      & 1.4     & 1.6     & 0.1     &  \B       \\

%
\hline 
\hline 
\multicolumn{7}{|c|}{} \\
\multicolumn{7}{|c|}{\mbox{\boldmath$g_V^{(I)}$}\bf-Perturbation Theory} \\
\hline
$\Lambda'_V$ [MeV]& \B       & \B       & 110(10) & 230(100)& \B       &
                                           \multicolumn{1}{|c|}{---} \\
$d^V_2$          & \B       & \B       & 0.05(5) & 0.7(6)  & \B       &  \B       \\
$\chi^2/dof$     & \B       & \B       & 1.4     & 1.6     & \B       &  \B       \\
\hline 
\hline 
\multicolumn{7}{|c|}{} \\
\multicolumn{7}{|c|}{\mbox{\boldmath$g_V^{(II)}$}\bf-Perturbation Theory} \\
\hline
$\Lambda'_V$ [MeV]& 147(4)  & 122(3)  & \B       & 120(20) & 140(70) & --- \\
$d^V_2$          & 0.20(2) & 0.01(1) & \B       & 0.01(8) & 0.1(3)  & 0.15(1) \\
$\chi^2/dof$     & 4       & 14      & \B       & 1.6     & 0.03    & 4 \\
%
\hline 
\multicolumn{7}{|c|}{} \\
\multicolumn{7}{|c|}{\mbox{\boldmath$g_{E2}$}\bf-Perturbation Theory} \\
\hline
$\Lambda_E$ [MeV]& \B       & 11.7(7) & \B       & 10(3)  & \B       & --- \\
$d^{E2}_2$       & \B       & 0.25(7) & \B       & 0.1(3) & \B       & 0.83(8)\\
$\chi^2/dof$     & \B       & 12      & \B       & 1.6    & \B       & 3 \\
%
\hline 
\hline 
\end{tabular}
\caption{ \it{Fits of lattice data using eq.(\ref{eq:gprime_fit})
(with an unknown 3-loop parameter $d_2$) with $g'$ defined as 
i) $g_{\overline MS}$,
ii) $g_V^{(I,II)}$, and
iii) $g_{E2}$.
A $\bullet$ signifies that no fit was found with $0 \le d_2 \le 1$.
}}
\label{tab:fit_3loop}
\end{center}
\end{table}

\begin{figure}[t]
\vspace{3.0truecm}

\ewxy{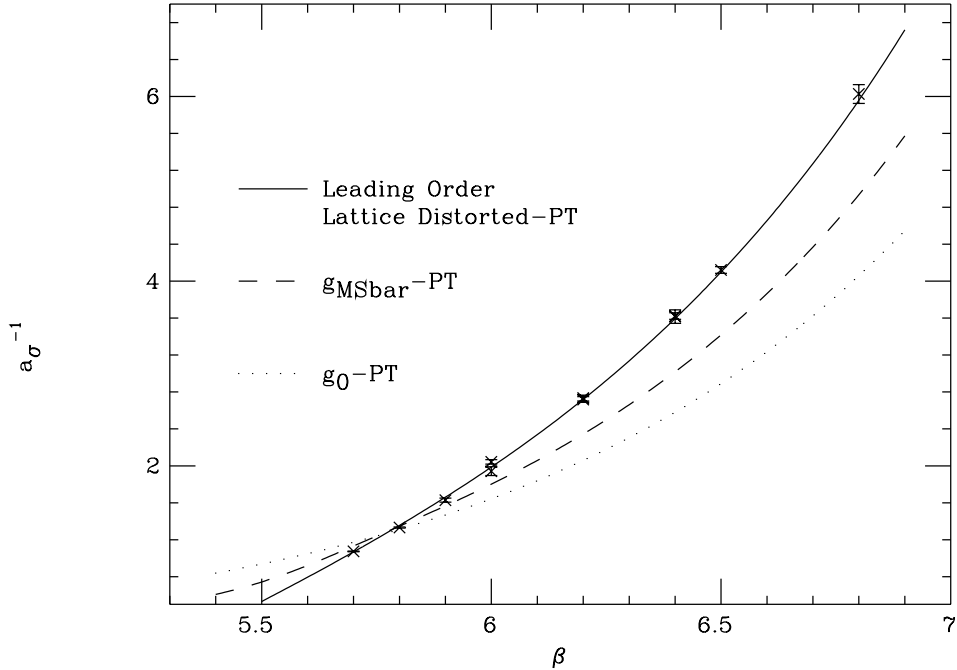}{85mm}

\vspace{2.5truecm}
\caption{\it Plot of $\ia$ from the string tension
together with fits from various methods described in the text.
In all cases the 2-loop function $f_{PT}$ was used.
The Monte Carlo data points are taken from table \ref{tab:inva}.
(The next-to-leading order lattice-distorted PT curve is not shown,
since it overlies the leading order lattice-distorted curve.)}
\vspace{1.0truecm}
\protect\label{fig:string}
\end{figure}

\begin{figure}[t]
\vspace{3.0truecm}

\ewxy{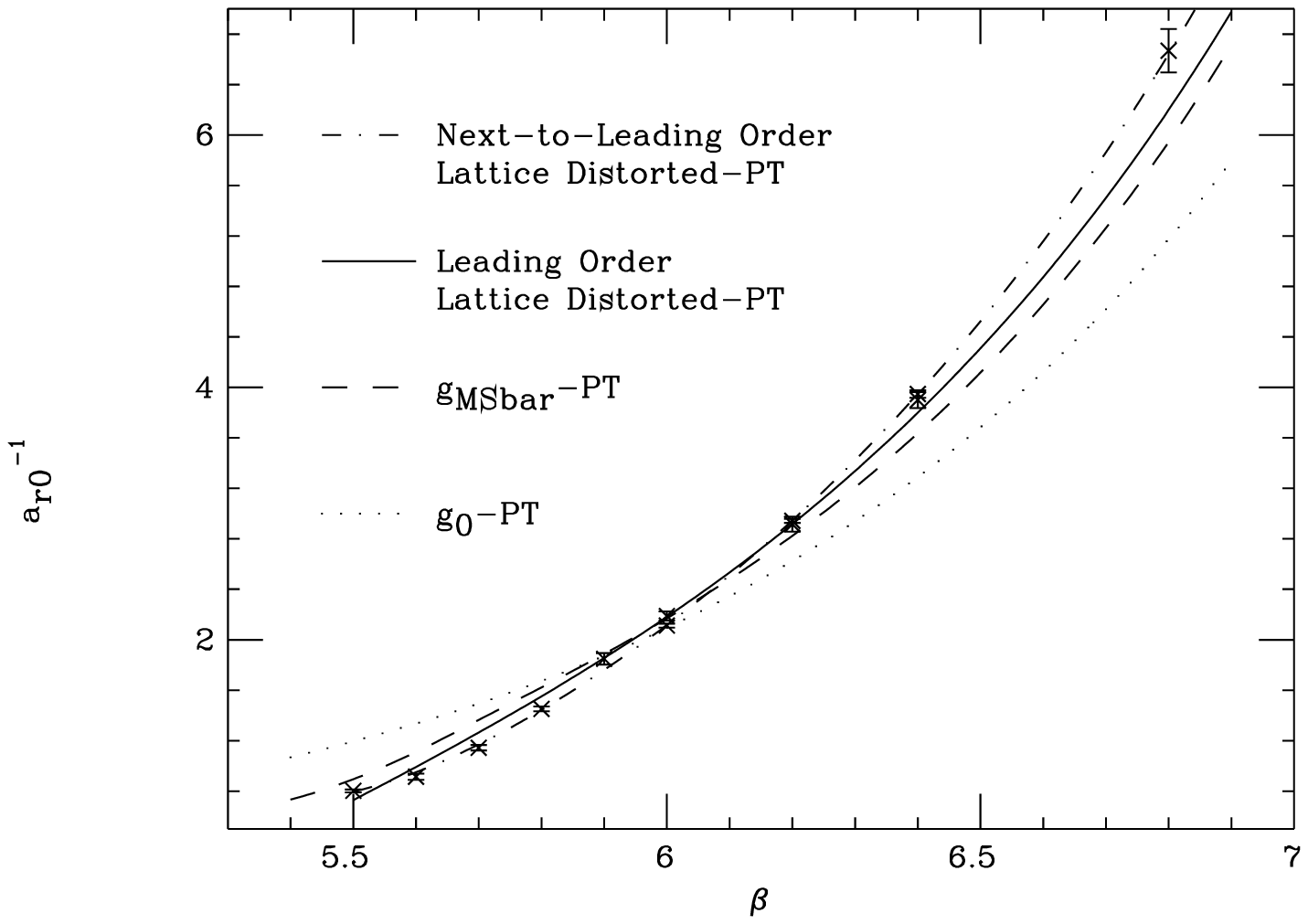}{85mm}

\vspace{2.5truecm}
\caption{\it Plot of $\ia$ from $r_0$
together with fits from various methods described in the text.
In all cases the 2-loop function $f_{PT}$ was used.
The Monte Carlo data points are taken from table \ref{tab:inva}.}
\vspace{1.0truecm}
\protect\label{fig:r0}
\end{figure}

\begin{figure}[t]
\vspace{3.0truecm}

\ewxy{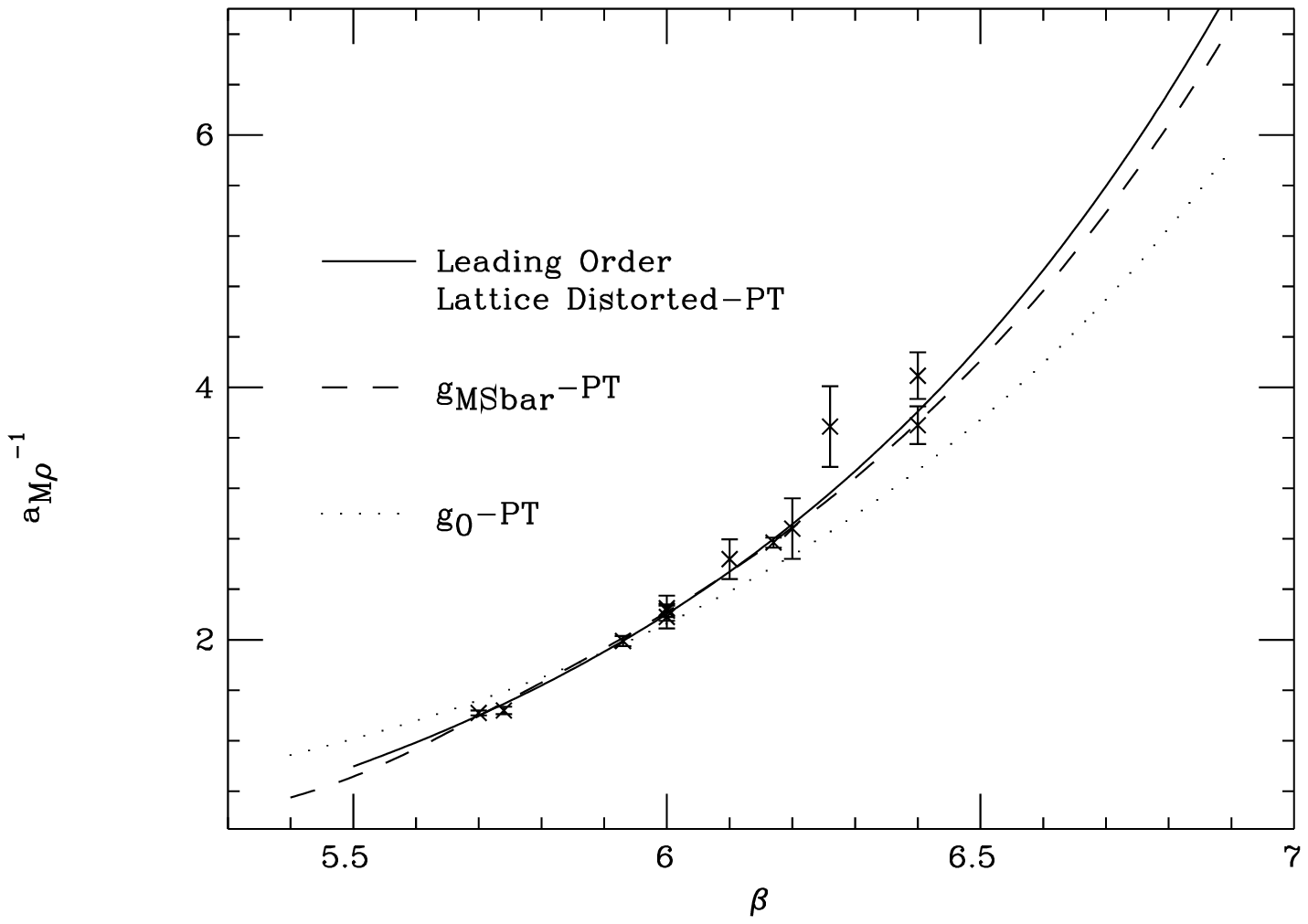}{85mm}

\vspace{2.5truecm}
\caption{\it Plot of $\ia$ from $M_\rho$
together with fits from various methods described in the text.
In all cases the 2-loop function $f_{PT}$ was used.
The Monte Carlo data points are taken from table \ref{tab:inva}.}
\vspace{1.0truecm}
\protect\label{fig:mrho}
\end{figure}

\begin{figure}
\vspace{3.0truecm}

\ewxy{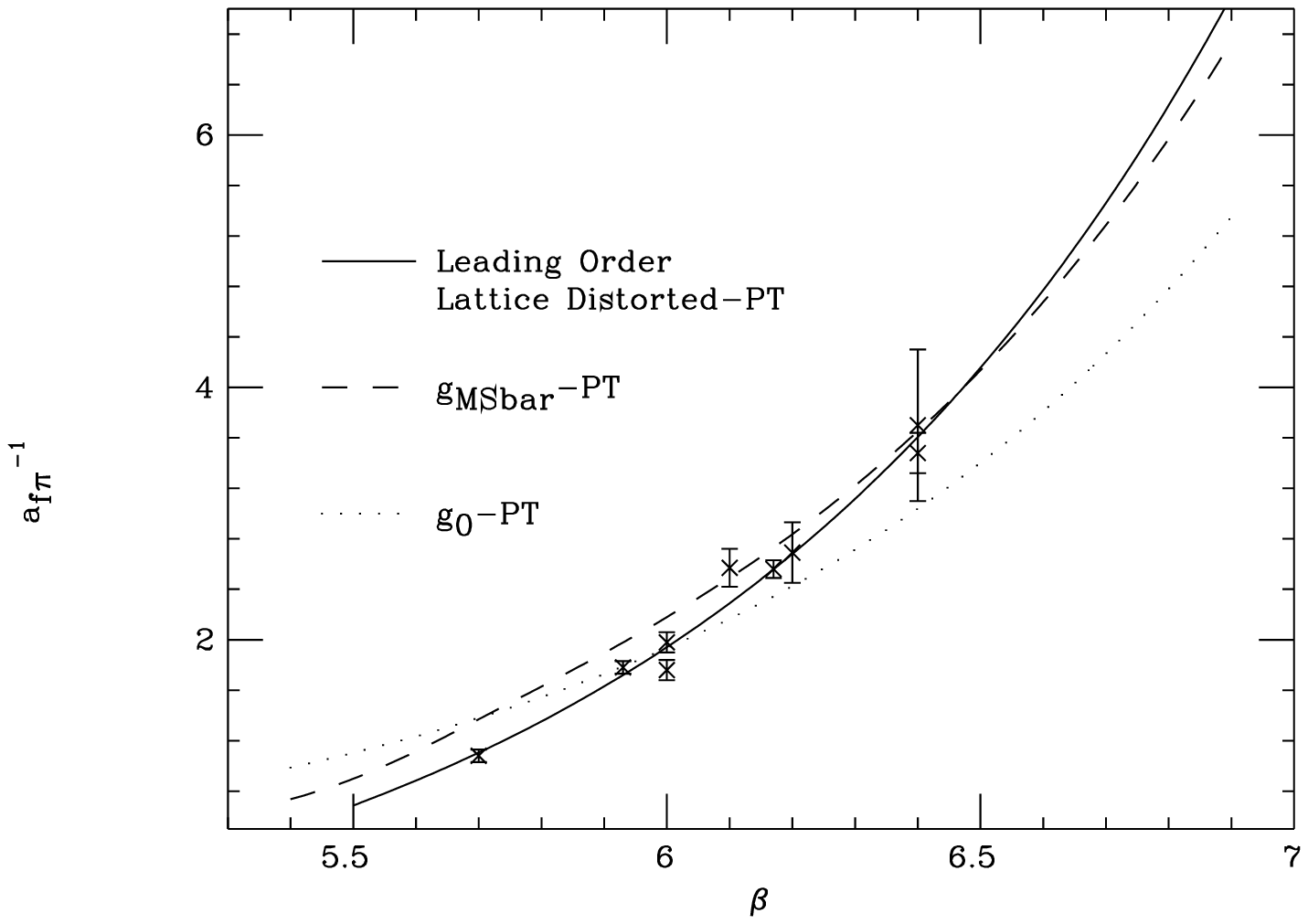}{85mm}

\vspace{2.5truecm}
\caption{\it Plot of $\ia$ from $f_\pi$
together with fits from various methods described in the text.
In all cases the 2-loop function $f_{PT}$ was used.
The Monte Carlo data points are taken from table \ref{tab:inva}.}
\vspace{1.0truecm}
\label{fig:fpi}
\end{figure}

\begin{figure}
\vspace{3.0truecm}

\ewxy{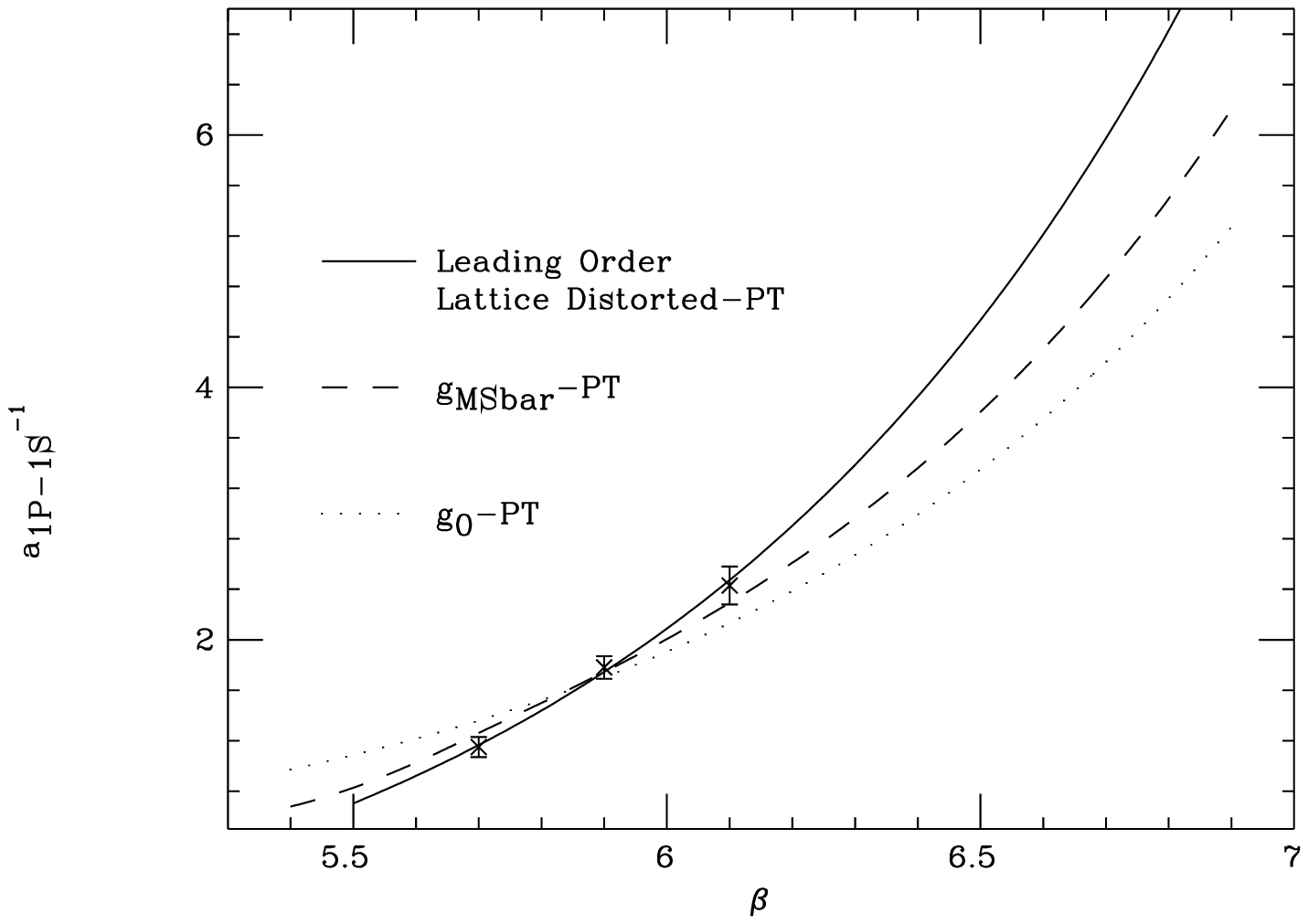}{85mm}

\vspace{2.5truecm}
\caption{\it Plot of $\ia$ from $1P-1S$ splitting
together with fits from various methods described in the text.
In all cases the 2-loop function $f_{PT}$ was used.
The Monte Carlo data points are taken from table \ref{tab:inva}.}
\vspace{1.0truecm}
\label{fig:1p1s}
\end{figure}

\begin{figure}
\vspace{3.0truecm}

\ewxy{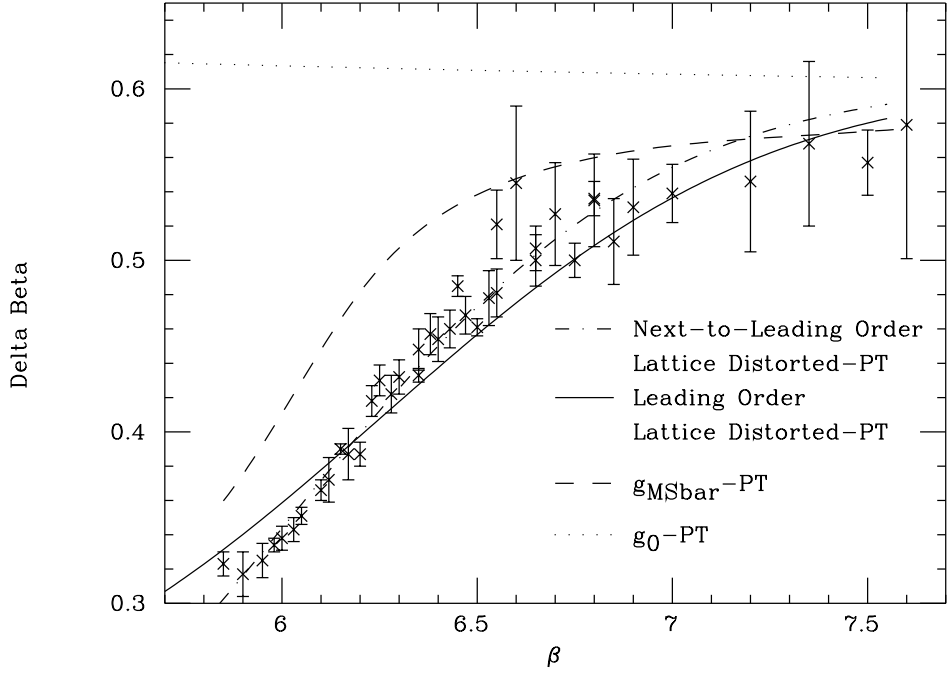}{85mm}

\vspace{2.5truecm}
\caption{\it Plot of $\Delta\beta(\beta)$
together with fits from various methods described in the text.
In all cases the 2-loop function $f_{PT}$ was used.
The Monte Carlo data points are from [25].}
\vspace{1.0truecm}
\label{fig:delta_beta}
\end{figure}

\begin{figure}

\setlength{\unitlength}{0.240900pt}
\ifx\plotpoint\undefined\newsavebox{\plotpoint}\fi
\sbox{\plotpoint}{\rule[-0.200pt]{0.400pt}{0.400pt}}%
\special{em:linewidth 0.4pt}%
\begin{picture}(1500,900)(0,0)
\font\gnuplot=cmr10 at 10pt
\gnuplot
\put(220,113){\special{em:moveto}}
\put(220,877){\special{em:lineto}}
\put(220,160){\special{em:moveto}}
\put(240,160){\special{em:lineto}}
\put(1436,160){\special{em:moveto}}
\put(1416,160){\special{em:lineto}}
\put(198,160){\makebox(0,0)[r]{0.6}}
\put(220,241){\special{em:moveto}}
\put(240,241){\special{em:lineto}}
\put(1436,241){\special{em:moveto}}
\put(1416,241){\special{em:lineto}}
\put(198,241){\makebox(0,0)[r]{0.65}}
\put(220,321){\special{em:moveto}}
\put(240,321){\special{em:lineto}}
\put(1436,321){\special{em:moveto}}
\put(1416,321){\special{em:lineto}}
\put(198,321){\makebox(0,0)[r]{0.7}}
\put(220,402){\special{em:moveto}}
\put(240,402){\special{em:lineto}}
\put(1436,402){\special{em:moveto}}
\put(1416,402){\special{em:lineto}}
\put(198,402){\makebox(0,0)[r]{0.75}}
\put(220,483){\special{em:moveto}}
\put(240,483){\special{em:lineto}}
\put(1436,483){\special{em:moveto}}
\put(1416,483){\special{em:lineto}}
\put(198,483){\makebox(0,0)[r]{0.8}}
\put(220,564){\special{em:moveto}}
\put(240,564){\special{em:lineto}}
\put(1436,564){\special{em:moveto}}
\put(1416,564){\special{em:lineto}}
\put(198,564){\makebox(0,0)[r]{0.85}}
\put(220,644){\special{em:moveto}}
\put(240,644){\special{em:lineto}}
\put(1436,644){\special{em:moveto}}
\put(1416,644){\special{em:lineto}}
\put(198,644){\makebox(0,0)[r]{0.9}}
\put(220,725){\special{em:moveto}}
\put(240,725){\special{em:lineto}}
\put(1436,725){\special{em:moveto}}
\put(1416,725){\special{em:lineto}}
\put(198,725){\makebox(0,0)[r]{0.95}}
\put(220,806){\special{em:moveto}}
\put(240,806){\special{em:lineto}}
\put(1436,806){\special{em:moveto}}
\put(1416,806){\special{em:lineto}}
\put(198,806){\makebox(0,0)[r]{1}}
\put(220,113){\special{em:moveto}}
\put(220,133){\special{em:lineto}}
\put(220,877){\special{em:moveto}}
\put(220,857){\special{em:lineto}}
\put(220,68){\makebox(0,0){0}}
\put(395,113){\special{em:moveto}}
\put(395,133){\special{em:lineto}}
\put(395,877){\special{em:moveto}}
\put(395,857){\special{em:lineto}}
\put(395,68){\makebox(0,0){0.05}}
\put(570,113){\special{em:moveto}}
\put(570,133){\special{em:lineto}}
\put(570,877){\special{em:moveto}}
\put(570,857){\special{em:lineto}}
\put(570,68){\makebox(0,0){0.1}}
\put(745,113){\special{em:moveto}}
\put(745,133){\special{em:lineto}}
\put(745,877){\special{em:moveto}}
\put(745,857){\special{em:lineto}}
\put(745,68){\makebox(0,0){0.15}}
\put(920,113){\special{em:moveto}}
\put(920,133){\special{em:lineto}}
\put(920,877){\special{em:moveto}}
\put(920,857){\special{em:lineto}}
\put(920,68){\makebox(0,0){0.2}}
\put(1095,113){\special{em:moveto}}
\put(1095,133){\special{em:lineto}}
\put(1095,877){\special{em:moveto}}
\put(1095,857){\special{em:lineto}}
\put(1095,68){\makebox(0,0){0.25}}
\put(1270,113){\special{em:moveto}}
\put(1270,133){\special{em:lineto}}
\put(1270,877){\special{em:moveto}}
\put(1270,857){\special{em:lineto}}
\put(1270,68){\makebox(0,0){0.3}}
\put(220,113){\special{em:moveto}}
\put(1436,113){\special{em:lineto}}
\put(1436,877){\special{em:lineto}}
\put(220,877){\special{em:lineto}}
\put(220,113){\special{em:lineto}}
\put( 5,495){\makebox(0,0){$a^{-1}_\sigma / a^{-1}_{PT}$}}                 
\put(828, 3){\makebox(0,0){$\{a_{PT}\}^2$}}
\put(1323,164){\raisebox{-.8pt}{\makebox(0,0){$\Diamond$}}}
\put(1101,271){\raisebox{-.8pt}{\makebox(0,0){$\Diamond$}}}
\put(924,370){\raisebox{-.8pt}{\makebox(0,0){$\Diamond$}}}
\put(782,448){\raisebox{-.8pt}{\makebox(0,0){$\Diamond$}}}
\put(578,595){\raisebox{-.8pt}{\makebox(0,0){$\Diamond$}}}
\put(448,684){\raisebox{-.8pt}{\makebox(0,0){$\Diamond$}}}
\put(312,770){\raisebox{-.8pt}{\makebox(0,0){$\Diamond$}}}
\put(402,708){\raisebox{-.8pt}{\makebox(0,0){$\Diamond$}}}
\put(782,513){\raisebox{-.8pt}{\makebox(0,0){$\Diamond$}}}
\put(578,603){\raisebox{-.8pt}{\makebox(0,0){$\Diamond$}}}
\put(448,681){\raisebox{-.8pt}{\makebox(0,0){$\Diamond$}}}
\put(1323,161){\special{em:moveto}}
\put(1323,167){\special{em:lineto}}
\put(1313,161){\special{em:moveto}}
\put(1333,161){\special{em:lineto}}
\put(1313,167){\special{em:moveto}}
\put(1333,167){\special{em:lineto}}
\put(1101,266){\special{em:moveto}}
\put(1101,276){\special{em:lineto}}
\put(1091,266){\special{em:moveto}}
\put(1111,266){\special{em:lineto}}
\put(1091,276){\special{em:moveto}}
\put(1111,276){\special{em:lineto}}
\put(924,354){\special{em:moveto}}
\put(924,387){\special{em:lineto}}
\put(914,354){\special{em:moveto}}
\put(934,354){\special{em:lineto}}
\put(914,387){\special{em:moveto}}
\put(934,387){\special{em:lineto}}
\put(782,418){\special{em:moveto}}
\put(782,479){\special{em:lineto}}
\put(772,418){\special{em:moveto}}
\put(792,418){\special{em:lineto}}
\put(772,479){\special{em:moveto}}
\put(792,479){\special{em:lineto}}
\put(578,579){\special{em:moveto}}
\put(578,611){\special{em:lineto}}
\put(568,579){\special{em:moveto}}
\put(588,579){\special{em:lineto}}
\put(568,611){\special{em:moveto}}
\put(588,611){\special{em:lineto}}
\put(448,669){\special{em:moveto}}
\put(448,699){\special{em:lineto}}
\put(438,669){\special{em:moveto}}
\put(458,669){\special{em:lineto}}
\put(438,699){\special{em:moveto}}
\put(458,699){\special{em:lineto}}
\put(312,743){\special{em:moveto}}
\put(312,797){\special{em:lineto}}
\put(302,743){\special{em:moveto}}
\put(322,743){\special{em:lineto}}
\put(302,797){\special{em:moveto}}
\put(322,797){\special{em:lineto}}
\put(402,694){\special{em:moveto}}
\put(402,721){\special{em:lineto}}
\put(392,694){\special{em:moveto}}
\put(412,694){\special{em:lineto}}
\put(392,721){\special{em:moveto}}
\put(412,721){\special{em:lineto}}
\put(782,497){\special{em:moveto}}
\put(782,529){\special{em:lineto}}
\put(772,497){\special{em:moveto}}
\put(792,497){\special{em:lineto}}
\put(772,529){\special{em:moveto}}
\put(792,529){\special{em:lineto}}
\put(578,587){\special{em:moveto}}
\put(578,620){\special{em:lineto}}
\put(568,587){\special{em:moveto}}
\put(588,587){\special{em:lineto}}
\put(568,620){\special{em:moveto}}
\put(588,620){\special{em:lineto}}
\put(448,651){\special{em:moveto}}
\put(448,712){\special{em:lineto}}
\put(438,651){\special{em:moveto}}
\put(458,651){\special{em:lineto}}
\put(438,712){\special{em:moveto}}
\put(458,712){\special{em:lineto}}
\put(220,806){\usebox{\plotpoint}}
\multiput(220,806)(20.756,0.000){59}{\usebox{\plotpoint}}
\put(1436,806){\usebox{\plotpoint}}
\sbox{\plotpoint}{\rule[-0.400pt]{0.800pt}{0.800pt}}%
\special{em:linewidth 0.8pt}%
\put(220,806){\special{em:moveto}}
\put(1403,113){\special{em:lineto}}
\end{picture}

\caption{\it Plot of $\ia / \ia_{PT}$ versus $a^2_{PT}$
for the string tension, where $\ia_{PT} = \Lambda_L /
f_{PT}(g_0^2)$. The value of $\Lambda_L = 5.85$ MeV from table
\ref{tab:2fit} was used and the 2-loop form of $f_{PT}$ was used to
define $a_{PT}$.
\label{fig:str_v_a2}}
\end{figure}
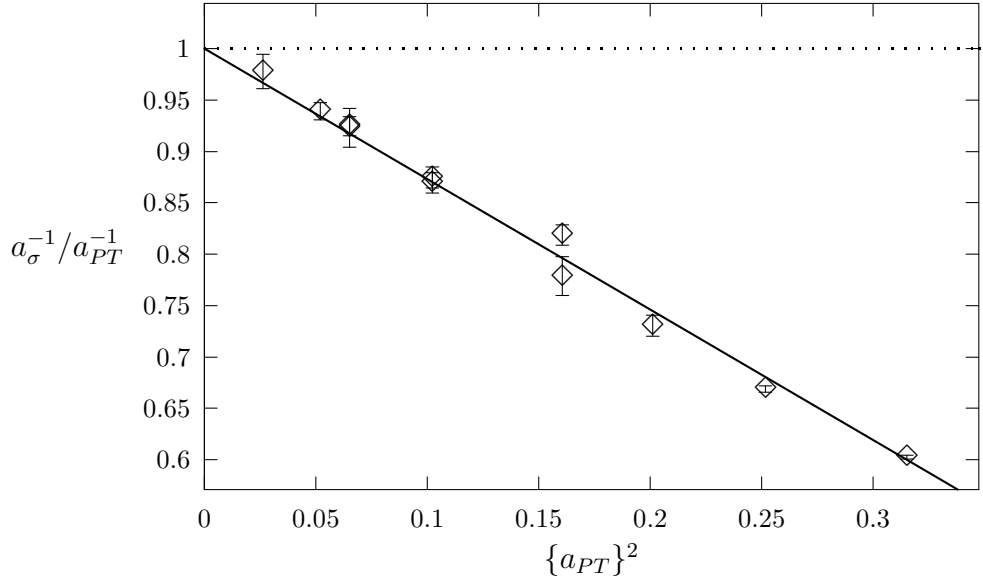

\section{Fits Using Lattice-Distorted PT}
\label{sec:ldpt_fit}

In this section we fit lattice Monte Carlo data for $\sigma, r_0
\cite{sommer}, M_\rho, f_\pi$ and the $1P-1S$ splitting to the form
eq.(\ref{eq:ia_ldpt}) with the coefficients $d^L$ set to zero
appropriately (i.e. we ignore higher order effects in
$g_0^2$). The data for these quantities is displayed in table
\ref{tab:inva} together with the references in column 1.

We perform two fits to eq.(\ref{eq:ia_ldpt}).
The first with $f_{PT}$ defined with $d^L_{l-1}=0$ for $l\ge3$
(i.e. including only 2-loop terms - see eq.(\ref{eq:f_PT})).
The second is with $f_{PT}$ defined with $d^L_{l-1}=0$ for $l\ge4$
(i.e. including the newly calculated 3-loop term, $b^L_2$ \cite{b_2}).

With the exception of the $1P-1S$ splitting \cite{aida} which comes
from the clover action \cite{sw,heatlie},
we consider results only from the Wilson action \cite{wilson}.
This is because there is not yet a great deal of data at a
wide range of $\beta$ values for non-Wilson actions.

We also fit the discrete beta function $\Delta\beta$
defined implicitly by
\be
a_L(\beta - \Delta\beta(\beta)) = 2 a_L(\beta)
\label{eq:delta_beta}
\ee
where $a_L$ is again defined from eq.(\ref{eq:ia_ldpt})
with the same constraints on the coefficients as mentioned above.
We use the $\Delta\beta$ data from the QCD-Taro collaboration
\cite{taro}.

In these fits we begin by including only the {\em leading}
${\cal O}(a^n)$ term appropriate for each quantity.
To be specific we rewrite eq.(\ref{eq:ia_ldpt}) as
\be
\ia_L(g_0^2) = \frac{\Lambda_L}{f_{PT}(g_0^2)}
\times  \left[ 1 - \x \frac{g_0^\nu f_{PT}^n(g_0^2)}
                       {           f_{PT}^n(1)    } \right],
\label{eq:ldpt_fit}
\ee
where there is no implicit summation over $n$ and $\nu$.
Note that the ${\cal O}(a^n)$ coefficient has been normalised so that $\x$
is the fractional amount of the ${\cal O}(g_0^\nu a^n)$ correction
at a standard value of $g_0=1$, corresponding of course to $\beta=6/g_0^2=6$.
Also note that we have approximated the polynomial coefficient, $c'_n(g_0^2)$,
by its leading ${\cal O}(g_0^2)$ term.
For each physical quantity listed above this leading term corresponds
to the following values of $n$ and $\nu$:
\bea\nonumber
M_\rho, f_\pi,1P-1S \;\;\; : \;\;\; & \nu = 0, n=1 \\\nonumber
\sigma, \Delta\beta \;\;\; : \;\;\; & \nu = 0, n=2 \\
r_0                 \;\;\; : \;\;\; & \nu = 2, n=2.
\label{eq:alphan}
\eea
The values in eq.(\ref{eq:alphan}) arises because different
quantities have different discretisation effects.
For example, $\sigma$, is measured using the gluonic
part of the lattice action only, and hence is correct to ${\cal O}(a^2)$.
The scale $r_0$ is correct to ${\cal O}(g_0^2 a^2)$
\cite{sommer}, hence the values $\nu=2, n=2$ for this quantity, etc.

A simple least squares fit of each column of data in table \ref{tab:inva}
to eq.(\ref{eq:ldpt_fit}) provides the values for $\Lambda_L$,
the coefficient $\x$, and the $\chi^2$ as listed in tables
\ref{tab:2fit} \& \ref{tab:3fit}
in the rows marked {\em Leading-Order Lattice Distorted PT}.
Table \ref{tab:2fit} contains fits using the 2-loop $f_{PT}$, and table
\ref{tab:3fit} has fits using the 3-loop function.\footnote{
Note that the fitting parameters for $\Delta\beta$ in table
\ref{tab:2fit} correct a slight error in the corresponding values in \cite{cra}.
}
In the case of $\Delta\beta$, we use eq.(\ref{eq:delta_beta}) with
$a_L$ defined in eq.(\ref{eq:ldpt_fit}).

Also shown in tables \ref{tab:2fit} \& \ref{tab:3fit} are the fits to
2-loop and 3-loop $g_0$-PT.\footnote{
The functional form of this fit is simply
$\ia_L(g_0^2) = \Lambda_L / f_{PT}(g_0^2)$
(i.e. eq.(\ref{eq:ldpt_fit}) with $\x\equiv0$).}
We see that leading order lattice-distorted PT fits the data very well
compared to the $g_0$-PT case, with the $\chi^2/dof$ down by up to two
orders of magnitude.
Figures \ref{fig:string}-\ref{fig:delta_beta} display the above fits in
the case of the 2-loop fit.

As a further check of the method, we include in the fit the next-to-leading
term in $a$.
However, due to the large statistical errors in some of the lattice
data we perform this fit only for $\sigma, r_0$ and $\Delta\beta$
where the statistical errors are very small.
The fitting function in these cases is:
\be\nonumber
\ia_L(g_0^2) = \frac{\Lambda_L}{f_{PT}(g_0^2)}
\times  \left[ 1 - \x   \frac{g_0^\nu f_{PT}^n(g_0^2)}
                        {           f_{PT}^n(1)    }
            - \xtwo\frac{g_0^\nu f_{PT}^{n+2}(g_0^2)}
                        {           f_{PT}^{\;n+2}(1)  } \right].
\label{eq:ldpt_fit2}
\ee
The results of these fits are again displayed in tables \ref{tab:2fit} \&
\ref{tab:3fit} for the 2-loop and 3-loop cases.
The figures \ref{fig:r0} and \ref{fig:delta_beta} also
display the fit in the 2-loop case for $r_0$ and $\Delta\beta$.

Obviously in the limit of infinite statistical precision,
reasonable fits to the lattice-distorted PT formula would only be
obtained if the ${\cal O}(a^n)$ terms were included to all orders.
The fact that it is necessary to go to next-to-leading order
for the $\sigma$, $r_0$ and $\Delta\beta$ data to obtain
a sensible $\chi^2/dof$ simply states that these data
have sufficiently small statistical errors to warrant this
order fit.
From here on, we take the results of the next-to-leading order fits to the
$\sigma$, $r_0$ and $\Delta\beta$ data as our best fits for these
quantities.

Some comments about the lattice-distorted PT procedure are necessary.

\begin{itemize}

\item The most important point is that for the $\sigma, r_0$ and
$\Delta\beta$ data, the agreement between the data and
lattice-distorted PT is truly remarkable considering the tiny
statistical errors in the lattice data.\footnote{
In fact, a very close look at the second order fit to $r_0$ for the
data from \cite{bali} resolves a small discrepancy between fit and data
for $\beta = 5.9$. It is possible that this is due to finite volume
effects since $L/a$ was $16$ for $\beta \le 5.9$ and $32$ for $\beta \ge
6.0$ for this data \cite{bali}.}

\item Another important finding is that the values of $\Lambda_L$ for
the various quantities in the 3-loop case are all consistent with
$\Lambda_L=7.7$ MeV within $1\sigma$. The only exception
is the string tension. Since the ``experimental'' value of the
string tension requires certain model assumptions, it is not possible
to draw any firm conclusion from this last observation.
While the effects of quenching will mean that, at some level, the various
values of $\Lambda_L$ will not all coincide, it is encouraging
that the $\Lambda_L$ values agree to $\sim 10\%$
\footnote{Because the simulations were all quenched, it would
be incorrect to perform a simultaneous fit to all the quantities
\{$\sigma,r_0,M_\rho,f_\pi,1P-1S$\} using a {\em single} $\Lambda_L$.}.
We take $\Lambda_L = 7.7 \pm 10\%$ MeV as an overall average where the
error includes an estimate (obtained by a comparison of the 2 and 3-loop
fits) of the effect of the unknown higher order terms $b^L_3$, $b^L_4$ etc.
Using $\Lambda_{\overline{MS}} = 28.81 \times \Lambda_L$, \cite{lambda_lat},
we have,
\[
\Lambda^{N_f=0}_{\overline{MS}} = 220 \pm 20 \;\;\;\mbox{MeV},
\]
where the superscript $N_f=0$ refers to the quenched approximation.
This compares very well with other lattice determinations of
$\Lambda^{N_f=0}_{\overline{MS}}$ \cite{lambda_MSbar} and therefore is
support for the validity of this approach.

\item The typical values of $\x$ in table \ref{tab:2fit} is 20-40\%.
In \cite{abada2}, it was found that
non-perturbative determinations of the renormalisation constant of
the local vector current, $Z^{Ren}_V$, vary by 10-20\%
depending on the matrix element used.
This spread in $Z^{Ren}_V$ has been interpreted as ${\cal O}(a)$
effects \cite{heatlie}.
Note that the simulation in \cite{abada2} was at $\beta=6.4$
where the value of $a$, and hence the size of the ${\cal O}(a)$ term,
is around half that at $\beta=6.0$.
Thus the values we obtain for $\x$ in this study are directly comparable
with the ${\cal O}(a)$ effects already uncovered in \cite{abada2},
confirming the lattice-distorted PT picture.

\item A compelling plot supporting lattice-distorted PT is shown in
figure \ref{fig:str_v_a2}. In this graph $a^{-1}_\sigma / a^{-1}_{PT}$
is plotted against $a^2_{PT}$, where $\ia_{PT}=\Lambda_L/f_{PT}(g_0^2)$.
The 2-loop form of $f_{PT}$ is used, however, in this plot, the 3-loop
would be indistinguishable.
The value of $\Lambda_L = 5.85$ MeV from table \ref{tab:2fit} is used.
If $g_0$-PT were correct (i.e. if there were perturbative scaling)
then the behaviour would be constant. Clear evidence for {\em linear}
behaviour with non-zero slope is apparent, implying that the
dominant corrections to $g_0$-PT are of ${\cal O}(a^2_{PT})$.

\item The coefficients for the second order terms, $\xtwo$ are an order
of magnitude smaller than the first order terms, $\x$. This follows
our expectation that the expansion in $f_{PT}$ in eq.(\ref{eq:ia_ldpt})
forms a convergent series, 
and that the bulk of the cut-off effects at present values of $\beta$
are due to the leading order term.

\item One of the most exciting features of the lattice-distorted PT
approach is that it can reproduce the behaviour of $\Delta\beta$
(see fig.\ref{fig:delta_beta}). 3-loop $g_0$-PT predicts
\be
\Delta\beta(\beta) = \delta_0 + \delta_1 g_0^2 + \delta^L_2 g_0^4
\label{eq:delta_beta_pt}
\ee
with $\delta_0 = 12 b_0 \log 2$, $\delta_1 = 12 b_1 \log 2$ and
$\delta^L_2 = 12 b^L_2 \log 2$.
The dotted curve in figure \ref{fig:delta_beta} with a near constant
value of around 0.6, shows the behaviour predicted from 2-loop $g_0$-PT.
(The 3-loop term, $\delta_2 g_0^4$ brings this curve down by only around 2\%.)
The large discrepancy between lattice Monte Carlo data
and 2-loop $g_0$-PT for $\Delta\beta$ has long been noted \cite{mcrg}.
The conventional reason for this ``dip'' is that it is a remnant of
the line of first order phase transitions in the fundamental-adjoint
coupling plane which ends near the fundamental axis in the
vicinity of this dip \cite{pole}.
However, the lattice-distorted PT can clearly solve this problem without
resorting to other explanations.

Note in \cite{taro}, a fit was performed to their $\Delta\beta$ data
using two free parameters.
The first was a 3-loop coefficient $\delta_2$ (see eq.(\ref{eq:delta_beta_pt})),
and the second involved defining a renormalised coupling, $g_u$, with
a shift $c_0$ relative to the $g_{\overline{MS}}$ coupling \cite{aida},
\bea\nonumber
\frac{6}{g_u^2} & = & \frac{6}{g_{\overline{MS}}^2} - c_0 \\\nonumber
                & = & \frac{6<U_{Plaq}>}{g_0^2} + 0.15 - c_0.
\eea
A good fit was obtained with a sensibly small 3-loop term,
$\delta_2/\delta_0 = -0.013(1)$, but with a huge shift $c_0 = 2.61(4)$
which is unphysical since it leads to a
divergent value of $g_u^2$ at $\beta=6/g_0^2 \approx 5$ \cite{taro}.
Thus this attempt at reconciling the $\Delta\beta$ data with PT
proved unsuccessful.

\item Comparing the values of $\x$ and $\xtwo$ in the 2 and 3-loop cases
(i.e. tables \ref{tab:2fit} \& \ref{tab:3fit}) we see that they are
compatible. In fact the main changes between the 2 and 3-loop cases
are in the values of $\Lambda_L$ which increases by around 15-20\%.
This variation
is consistent with the value of $d^L_2$ (see eq.(\ref{eq:f_PT})). In
other words, the term $d^L_2 g^2$ can be well approximated by
$d^L_2$. This again implies that the higher order terms in $g_0^2$ are
not important in the comparison between perturbation theory and current
Monte Carlo data (apart from a normalising effect on $\Lambda_L$).

\item The values of $\x$ and $\xtwo$ in the case of $\sigma$ are
compatible with those for $\Delta\beta$. This is a comforting feature,
since both quantities are, in effect, obtained from the study of large
Wilson loops \cite{taro}.

\item As far as the fit to $M_\rho, f_\pi$ and the $1P-1S$
splitting are concerned, the errors in the lattice data are
large enough to allow many functional forms. Thus these
data do not constrain the lattice-distorted PT fit
(or fits from other schemes).

\item The coefficients $c'_n$ of the ${\cal O}(a^n)$ terms in
eq.(\ref{eq:ia_ldpt}) are, strictly speaking, polynomial functions of
$g_0^2$. In the above fits, we have taken only the leading term in this
polynomial. This is consistent with the lattice-distorted PT approach,
since it assumes that the leading term of ${\cal O}(a^n)$ appearing in
eq.(\ref{eq:ia_ldpt}) dominate the ${\cal O}(g_0^{2l-4})$ terms.
Consistency within this picture therefore implies that the polynomials
$c'_n(g_0^2)$ can be replaced with their leading term in $g_0^2$.

\item The errors in the fitting parameters $\x$ are statistical only.
They do not include any estimate of the systematic error due to the
non-inclusion of the still higher order terms in $a$.
A more refined fitting procedure would be required to estimate these
errors. This will be left for future studies.

\item The renormalisation constant used in the determination of
$a^{-1}_{f_\pi}$ is $Z^{Ren}(g_0^2) = 1 - 0.132 g_0^2$ \cite{mz}
i.e. for consistency we have used the bare lattice coupling
as the expansion parameter, rather than a renormalised coupling.
However, as already mentioned, the statistical errors in the
data for $a^{-1}_{f_\pi}$ are too large to enable useful
conclusions to be drawn from these data alone.

\end{itemize}

\section{Fits Using a Renormalised Perturbation Theory}
\label{sec:g'}

\subsection{Introduction to the Renormalised Coupling fits}

In this section we fit the Monte Carlo data for $a^{-1}$ to the
following functional form:
\be
\ia_L(g_0^2) = \frac{\Lambda}{f_{PT}((g')^2)},
\label{eq:gprime_fit}
\ee
where $g'$ is some ``renormalised'' coupling which is in turn a function
of the bare coupling $g_0$.

Note that the philosophy behind these fits is that the failure of
asymptotic (perturbative) scaling is explained by higher order terms in
perturbation theory. The coupling $g'$ is defined with the aim of
creating perturbative series with improved convergence properties. This
is an orthogonal philosophy compared to the procedure outlined in
Sect.~\ref{sec:ldpt_fit} where finite lattice spacing errors are
assumed to cause the disagreement between Monte Carlo data and
perturbative scaling. (Note that the ${\cal O}(a^n)$ terms in
eqs.(\ref{eq:ldpt_fit} \& \ref{eq:ldpt_fit2}) cannot be expressed as
polynomials in $g_0^2$ - i.e. they cannot be written in the form of the
$d^L_{l-1}g_0^{2l-4}$ terms in eq.(\ref{eq:ia_ldpt}).)

The following subsections study fits using various definitions of $g'$.
Both 2 and 3-loop fits were performed, analogous to the lattice
distorted case.
At the end of this section we make some general comments about the
success or otherwise of the renormalised coupling approach.

\subsection{Fits Using $g_{\overline{MS}}$-Perturbation Theory}
\label{sec:gms}

Following ref. \cite{aida} we define an ${\overline{MS}}-$like coupling, where
$g' = g_{\overline{MS}}$,
\[
\frac{1}{g_{\overline{MS}}^2(\pi/a)} = \frac{1}{g_0^2} \;
           <\frac{1}{3} Tr U_{plaq}>_{MC} + 0.025.
\]

The results of using this definition of $g'$ in the fitting function
Eq.(\ref{eq:gprime_fit}) (with the 2-loop definition of $f_{PT}$) are
displayed in Figs \ref{fig:string} - \ref{fig:delta_beta} and in Table
\ref{tab:2fit} in the rows headed $g_{\overline{MS}}$.

In order to perform 3-loop fits, an extra fitting parameter,
$d^{\overline{MS}}_2$, in $f_{PT}$ is required. This is because $d^L_2$
from \cite{b_2} is only appropriate for the $g_0$ scheme.
The results of these fits
are displayed in Table \ref{tab:fit_3loop}.

\subsection{Fits Using $g_V$-Perturbation Theory}
\label{sec:gv}

We now turn to some definitions of $g'$ proposed by Lepage and Mackenzie
\cite{lm}.
Ref. \cite{lm} lists three alternative ``renormalised'' couplings
$g_V^{(i)}$ ($i=I,II,III$) which are based on, or related to the strength
of the static quark anti-quark potential. They are:
\[
\alpha_V^{(I)}(\pi/a) = \frac{\alpha_L}{<\frac{1}{3} Tr U_{plaq}>_{MC}} \times
( 1 + 0.513 \alpha_V ),
\]
(see \cite{lm} eq(29)); and
\[
- \ln {<\frac{1}{3} Tr U_{plaq}>_{MC}} = \frac{4\pi}{3} \alpha_V^{(II)}(3.41/a) \times
( 1 - 1.19 \alpha_V ),
\]
(see \cite{lm} eq(20)); and
\[
\alpha_V^{(III)}(46.08/a) = \alpha_L,
\]
(see \cite{lm} eq(30))
where $\alpha \equiv g^2/4\pi$ throughout and $MC$ stands for Monte
Carlo estimate. Lepage
and Mackenzie  argue that all three definitions agree up to
${\cal O}(\alpha_V^3)$ (see discussion surrounding Fig. 7 of Ref. \cite{lm}).

Before checking the fits of $a^{-1}$ using these definitions of $g_V$,
we briefly discuss the effects of the momentum scale in the above
definitions. Lepage and Mackenzie argue that there is a momentum scale
$q^\ast$ which is most appropriate for each quantity studied which can
be obtained from a mean field calculation.
Thus, for example, the $q^\ast$ value for the critical quark mass for the
Wilson action is $q^\ast = 2.58/a$.
In order to fit the Monte Carlo data for $a^{-1}$ to 
Eq.~(\ref{eq:gprime_fit}) (with $g' = g_V$), we should, in theory, first
obtain a value for $q^\ast$ appropriate for each of the quantities
$\Omega$ studied
($\Omega = \sigma, r_0, M_\rho, f_\pi, $ or the $1P-1S$ splitting).
However, as is shown below, fits using Eq.~(\ref{eq:gprime_fit})
are not dependent on the scale $q^\ast$ chosen, so long as the parameter
$\Lambda_V$ is trivially rescaled.

We prove this statement as follows.
Since $g_V$ is a function of $q$ and $\beta$, we write $g_V(q,\beta)$ to
make these dependencies explicit. In order to calculate the coupling,
$g_V(q^\ast,\beta)$, at some new scale, $q^\ast$, we first must
determine $\Lambda_V$. This can be done using
\be
\Lambda_V  = q \; f_{PT}(g_V^2(q,\beta)).
\label{eq:lam}
\ee
Once $\Lambda_V$ has been determined, the value of $g_V$ can
be determined at the new scale $q^\ast$ using the implicit definition:
\be
\Lambda_V  = q^\ast \; f_{PT}(g_V^2(q^\ast,\beta)).
\label{eq:qstar}
\ee

Now we turn to the fit of the $a^{-1}_L$ obtained from a Monte Carlo
determination of some physical quantity $\Omega$. The appropriate fitting
function (see Eq.(\ref{eq:gprime_fit})) is
\[
\ia_L(g_0^2) = \frac{\Lambda_V} {f_{PT}(g^2_V(q^\ast,\beta))} \;.
\]
Using Eqs.(\ref{eq:lam} \& \ref{eq:qstar}) we immediately have
\be
\ia_L(g_0^2) = \frac{q^\ast/q \; \Lambda_V}
                    {f_{PT}(g^2_V(q,\beta))}
             = \frac{\Lambda'_V(q)} {f_{PT}(g^2_V(q,\beta))} \;,
\label{eq:g_V_fit}
\ee
where $\Lambda'_V(q) \equiv \frac{q^\ast}{q} \Lambda_V$. So the fit
is independent of the choice of $q^\ast$ apart from a trivial rescaling
of $\Lambda_V$, as claimed. Therefore we do not need to worry about the
choice of $q^\ast$ for the quantities $\Omega$ considered in this
study. Note, however, that for quantities such as the plaquette which
perturbatively are expressed as a {\em polynomials} in $g^2$ rather than
proportional to $f_{PT}$, the above argument does not hold, and
results {\em do} depend on the choice of $q^\ast$. This last fact has
been shown in Ref. \cite{lm}.

The results of the fits of the $a^{-1}$ data to the functional form
in Eq.~(\ref{eq:g_V_fit}) (with the 2-loop definition of $f_{PT}$)
are shown in Table \ref{tab:2fit} in the rows headed $g_V^{(I,II)}$. 
The coupling $g_V^{(III)}$ was not considered since it would be the lead
to the same fit as the $g_0$ case apart from a trivial constant factor
in the $\Lambda_V$ value as discussed above.
The $q$ values for the fits $I,II$ are $q=\pi/a,\,3.41/a$
respectively.
Since it is only the quality of the fits that is of
interest no attempt was made to run the coupling $g_V^{(I,II)}$ to the
same momentum scale.
However if this was attempted, inconsistent results would be obtained,
since the $\Lambda'_V(q)$ values do not appear to fit the relationship
$\Lambda'_V(q^{(I)}) \equiv \frac{q^{(II)}}{q^{(I)}} \Lambda'_V(q^{(II)})$.

As in the ${\overline{MS}}$ case, in order to perform 3-loop fits, an
extra fitting parameter, $d^V_2$, in $f_{PT}$ is required.
The results of these fits are displayed in Table \ref{tab:fit_3loop}.

\subsection{Fits Using $g_E$-Perturbation Theory}
\label{sec:ge}

Following refs. \cite{parisi} and \cite{wup1} we define a coupling,
$g_E$, based on the plaquette,
\[
\frac{1}{g_E^2} = \frac{c_1}{1 - <\frac{1}{3} Tr U_{plaq}>_{MC}}.
\]
where $c_1 = \frac{1}{3}$ is the coefficient of the first term in the
perturbative expansion for $1 - <\frac{1}{3} Tr U_{plaq}>$.
The results using this definition for $g'$ in the fitting function
eq.(\ref{eq:gprime_fit}) are displayed in Table \ref{tab:2fit} in the
rows headed $g_E$-PT for the case of the 2-loop fits.
For the 3-loop fit, the coefficient appropriate for the $g_E$ scheme
can be derived from a short calculation \cite{b_2}, $d^E_2=0.01161$.
The results for these fits are displayed in Table \ref{tab:3fit}.

In ref \cite{wup1}, the $g_E$ scheme was extended to the next
order in the perturbation series. That is, a coupling $g_{E2}$, is
defined by demanding that the Monte Carlo result for the plaquette
is equal to the perturbative series truncated to second order. i.e.
\[
1 - <\frac{1}{3} Tr U_{plaq}>_{MC} = c_1 g_{E2}^2 + c_2 g_{E2}^4,
\]
where $c_2 = 2 \times (0.0204277 - 1/288)$ \cite{c_2}.
The results of this fit are displayed in Table \ref{tab:2fit} in the
rows headed $g_{E2}$ for the 2-loop fit.
As in the ${\overline{MS}}$ case, in order to perform 3-loop fits, an
extra fitting parameter, $d^{E2}_2$, in $f_{PT}$ is required.
The results of these fits are displayed in Table \ref{tab:fit_3loop}.

\subsection{Discussion of the Renormalised Coupling fits}

We begin with a discussion on the quality of the fits in the 2-loop case.
As can be seen from Table~\ref{tab:2fit}, the fits using $g_{\overline{MS}}$,
$g_V^I$, $g_V^{II}$, $g_E$ and $g_{E2}$ -PT are not as successful as
those from lattice distorted-PT.
It is interesting to note that of the two $g_V$ fits, the $g_V^{II}$
definition seems more successful.
Surprisingly, the $g_{E2}$ fit is not a great deal better than the $g_E$
fit, even though it should naively be correct to one more order of
perturbation theory.
This in itself suggests that pushing to higher orders in perturbation
theory does not pay significant dividends.

Concentrating on the fits to $\sigma$, $r_0$ and $\Delta\beta$ where the
lattice data has very small statistical errors, we see from
Table~\ref{tab:2fit} that the $g_V^{II}$ and the $g_{E2}$ -PT methods are
the most successful of the renormalised coupling approaches.
In fact the $g_V^{II}$ and $g_{E2}$ -PT fits to $r_0$ are comparable in
quality to the leading-order lattice distorted PT.
However, these same $g_V^{II}$ and $g_{E2}$ fits are an order of magnitude
worse for the $\sigma$ case compared to the leading-order lattice
distorted PT, and are also significantly worse for the fits to $\Delta\beta$.

Of course the (leading order) lattice distorted PT fits include one
extra fitting parameter ($\x$) compared to the renormalised coupling
fits, so it is tempting to argue that the former fits are more
successful only for this reason.
However, Sec.\ref{sec:ldpt_fit} details the results from a renormalised coupling
style fit with {\em two} extra fitting parameters that proved entirely
unsuccessful \cite{taro}.

Turning to the 3-loop case, even more striking comments can be made.
While it remains true that the lattice distorted PT fits remain
stable and of high quality, the same cannot be said for the renormalised
coupling approach. The $g_E$ scheme (whose 3-loop coefficient is known)
cannot reproduce the quality of fits of the (even) the leading order
lattice distorted PT (see Table \ref{tab:3fit}).
Furthermore, of the four schemes listed in Table
\ref{tab:fit_3loop}, 
none can provide sensible fits for all the quantities studied.
A ``sensible fit'' is defined, very liberally, as one where the fitted
value for $d_2$ lies in the range $0 \le d_2 \le 1$.
Note also that in the cases where sensible fits are obtained,
the values of $d_2$ depend strongly on the quantity being studied - a clear
indication that there are non-universal (finite-lattice spacing) effects
present. Furthermore, the values of the coefficient $d_2$ can often be
$\sim 20\%$ or more. According to the renormalised coupling philosophy,
$d_2$ should be significantly smaller
than the corresponding coefficient in the $g_0$ scheme. However, since $d^L_2
\sim 20 \%$, this is clearly not the case.

In conclusion none of the renormalised coupling methods approach
the quality and consistency of the fits from lattice distorted PT method.

\section{Discussion \& Conclusions}

In this paper we discuss the possible causes of the lack of
perturbative scaling in dimensionful lattice quantities.
We have argued that the cause of this disagreement is due to either
(i) the neglect of higher order terms in perturbation theory (i.e. the
renormalised coupling approach), and/or
(ii) the presence of cut-off effects due to the finiteness of the lattice
spacing, $a$ (i.e. the lattice distorted PT approach).
Taking a collection of quenched lattice data from various collaborations
for the string tension $\sigma$, the hadronic scale $r_0$, $M_\rho$, $f_\pi$,
the $1P-1S$ splitting in charmonium and the discrete beta function
$\Delta\beta$, we have attempted fits assuming each of the
hypotheses (i) and (ii) above.
We have found that modelling the discrepancy using (i),
through the use of a renormalised coupling, $g'$, provides less
satisfactory fits to the data compared to (ii).
In fact, the hypothesis (ii) leads to a remarkable consistency both in the
fitted $\Lambda$ parameters and amongst the sizes of the ${\cal O}(a^n)$
corrections.
Furthermore, the hypothesis (ii) succinctly explains the
``non-perturbative'' behaviour of $\Delta\beta$.
This, we feel, is one of its major achievements.
Using the fits from lattice distorted PT (method (ii)) we estimate
$\Lambda^{N_f=0}_{\overline{MS}} = 220 \pm 20$ MeV, perfectly
compatible with values obtained from other methods.
We find excellent consistency for this $\Lambda_{\overline{MS}}$ value amongst
the different physical quantities studied.

It is clear however, that we have not {\em proven} that the lack of
perturbative scaling is due to finite lattice spacing effects
(i.e. hypothesis (ii)).
In fact, presumably, the true state of affairs is that the lack of
perturbative scaling is due to a mixture of the two effects (i) and (ii).
However, we have given strong arguments to support the claim that the
discrepancy between the perturbative scaling formula and current Monte
Carlo data is dominated by lattice artifacts (see Sec.\ref{sec:ldpt_fit}).

\begin{figure}
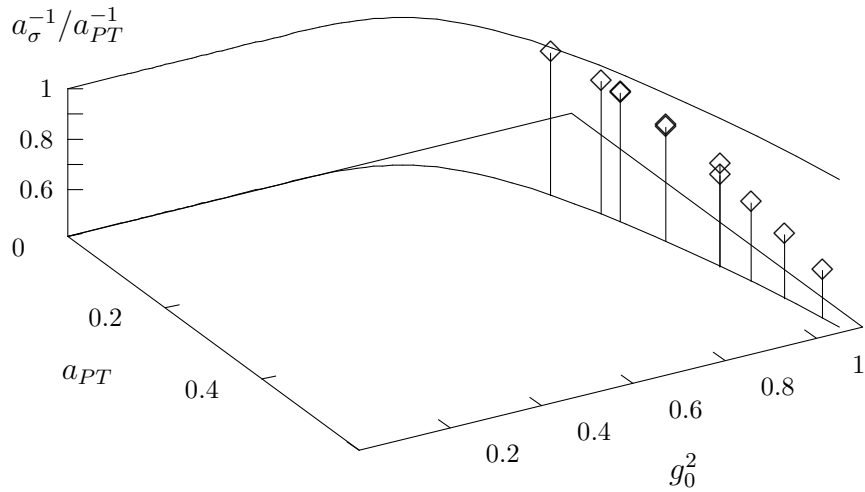


\input 3dim_string.tex

\caption{\it Plot of $\ia/\ia_{PT}$ versus $a_{PT}$ and $g_0^2$.
The string tension is used to set $a$.
$a_{PT}$ is defined as $\Lambda_L / f_{PT}(g_0^2)$
where $\Lambda_L = 5.85$ MeV is taken from Table~\ref{tab:2fit}
and the 2-loop version of $f_{PT}$ is used.
The data points are plotted using diamonds (the error bars are suppressed
since they are smaller than the symbols).
The vertical lines joining them to the $a_{PT}-g_0^2$ axis are guides
for the eye only.
The curves in the $a_{PT}-g_0^2$ plane are $a_{PT} = f_{PT}(g_0^2) / \Lambda_L$
with $\Lambda_L$ defined as above.
(No discernible difference in this plot would be observed if the 3-loop
version of $f_{PT}$ were used instead.)
\label{fig:3dim_string}}
\end{figure}

A pictorial representation of the approach to perturbative scaling in
lattice data is displayed in  Figure~\ref{fig:3dim_string}.
Here the ratio $\ia_\sigma / \ia_{PT}$ is plotted on a 3-dimensional graph
against both $g_0^2$ and $a_{PT}$.
$\ia_\sigma$ is defined from the Monte Carlo values for the string
tension (from Table \ref{tab:inva}),
and $a_{PT}$ is defined, $\ia_{PT}(g_0^2) = \Lambda_L / f_{PT}(g_0^2)$,
where $\Lambda_L = 5.85$ MeV is taken from Table~\ref{tab:2fit}.
(In the actual plot the 2-loop formula for $f_{PT}$ has been used.
However, the 3-loop plot would be indistinguishable.)
We use the string tension in this graph since the lattice data
has very small statistical errors, however other quantities would lead
to similar plots.
The diamonds are the data points (error bars have been omitted because
they are smaller than the symbols),
and the vertical lines joining the data
points to the $a_{PT}-g_0^2$ plane are guides for the eye only.
The two curves shown are defined by
$a_{PT}(g_0^2) = f_{PT}(g_0^2) / \Lambda_L$,
with $\Lambda_L$ defined as above.
The lower curve is a guide for the eye, and the upper curve is
in the $\ia_\sigma / \ia_{PT} = 1$ plane.
This upper curve shows the relationship between the Monte Carlo derived
lattice spacing, $a$, the bare coupling, $g_0^2$, and the perturbative
formula for the lattice spacing, $a_{PT}$ in the idealised case of
perfect (2-loop) perturbative scaling.
It is important to note that the deviation between the data points
and this curve in Figure~\ref{fig:3dim_string}
is precisely the discrepancy being studied in this paper.
\footnote{Note that Figure~\ref{fig:str_v_a2} is a projection of 
Figure~\ref{fig:3dim_string} onto the $\ia_{\sigma} /
\ia_{PT}-a_{PT}$ plane.
However the data in Figure~\ref{fig:str_v_a2} is plotted versus
$a_{PT}^2$ rather than $a_{PT}$.}

The ratio $\ia_\sigma/\ia_{PT}$ is interesting since from eq.(\ref{eq:ia_ldpt})
it is unity up to terms ${\cal O}(g_0^2)$ and ${\cal O}(a_{PT})$, i.e.

\be
\frac{\ia_\sigma}{\ia_{PT}} = ( 1 + \sum_{l=4} d^L_{l-1} g_0^{2l-4} )^{-1} \times
                       ( 1 + \sum_{n=1} c'_n(g_0^2) a_{PT}^n(g_0^2) ).
\label{eq:ia:iapt}
\ee

From Figure~\ref{fig:3dim_string} the ratio $\ia_{\sigma}/\ia_{PT}$
varies from 0.6 to around 1.0 over the range of presently available
lattice data.
It can also be seen that the range in $a_{PT}$ available from current lattice
data spans a factor of 3-4,
whereas the corresponding range in $g_0^2$ is very small - from around
0.88 to 1.05 only (i.e. a spread of $\ltap 20\%$).
Thus, the best of the current data is relatively a long way from the
continuum limit $a=g_0^2=0$ in terms of variable $g_0^2$,
but apparently relatively close in terms of $a$.
Therefore current lattice data is in the regime where effects of ${\cal O}(a)$,
${\cal O}(a^2)$ etc, will be seen clearly (if these terms are significant)
since there is a broad range in $a$ available.
Conversely, higher order effects in $g_0^2$ will be much more difficult
to discern since the range available in this variable is so compact.

We now aim to reconcile the rapid variation of $\ia_{\sigma}/\ia_{PT}$
from 0.6 to around 1.0 with eq.(\ref{eq:ia:iapt}).
Considering the relative variations in $g_0^2$ and $a_{PT}$,
it is natural to assume that the ${\cal O}(a^n)$ term
(with $n=2$ for the string tension) is more dominant than the
${\cal O}(g_0^2)$ term.
Note that the fact that significant ${\cal O}(a)$ corrections have
already been uncovered in lattice data adds support to these arguments
(see Sec.\ref{sec:ldpt_fit}).

To be more quantitative, we have performed a fit of the form
\[
\frac{\ia_\sigma}{\Lambda_L/f_{PT}(g_0^2)} = ( 1 + d^L_3 g_0^4 )^{-1} \times
                                      ( 1 + c'_2 a_{PT}^2(g_0^2) ),
\]
to the $\sigma$ and $r_0$ data in table \ref{tab:inva}.
Obviously this fitting form allows
both an ${\cal O}(a^2)$ distortion, and a 4-loop correction.
The result is a worse fit than the next-to-leading order lattice
distorted fit (which has the same number of free fitting parameters)
in the case of the $\sigma$ data, and no sensible fit at all in the case
of the $r_0$ data.

Of course this argument again does not {\em prove} the case for lattice
distorted PT; one could rely on the conspiracy of higher order
terms in perturbation theory to recover the behaviour seen in
Figure~\ref{fig:3dim_string} without the need for any ${\cal O}(a^n)$
terms at all.
However, it seems much more natural, taking into account the arguments
outlined above and in Sec.\ref{sec:ldpt_fit}, to rely on the
${\cal O}(a^n)$ terms to explain the discrepancy instead.

Using these assumptions, we can determine the value of the coupling when
the finite lattice spacing effects first drop below, say, 1\%.
Taking a typical value of $\x \approx 30 \%$ from
Table~\ref{tab:2fit} or \ref{tab:3fit},
the ${\cal O}(a^n)$ term in eq.(\ref{eq:ldpt_fit}),
$\approx X_n f_{PT}^n(g_0^2) / f_{PT}^n(1)$, is less than 1\% for $g_0^2 \ltap
0.7$, i.e. $\beta \gtap 9$ in the case of $n=1$ distorted quantities,
and for $g_0^2 \ltap 0.8$, i.e. $\beta \gtap 8$ for $n=2$ quantities.

Continuing the assumption that the discrepancy between lattice data and
perturbative scaling is due to finite lattice spacing effects, we can also
explain why it is that the renormalised coupling approach works as well
as it does.
This is presumably because the ${\cal O}(a)$ effects of the quantity
being used to define $g'$ are embedded in the definition of $g'$.
So long as the ${\cal O}(a)$ effects in the quantity being used to
define $g'$ are similar in sign and magnitude to those in the quantity
being studied (e.g. $\sigma$, $r_0$, etc.) the PT expressed in terms of
$g'$ will fit the lattice data better than the bare $g_0$ formula.
Thus the $g'$ definition mixes finite lattice spacing distortions with
higher order perturbative effects.

There is a final interesting opportunity assuming the correctness
of the lattice distorted perturbative scaling philosophy \cite{cra}.
Suppose one is studying the continuum limit of a dimensionful quantity,
$\Omega$, using lattice estimates determined at a set of finite $a$ values.
Typical examples of $\Omega$ are the pseudoscalar decay constant $f_\pi$
or the mass $M_\rho$.
The physical value of $\Omega$ can be written as

\be
\Omega = \lim_{g_0\rightarrow 0}\; [ Z^{Ren}(g_0^2) \times
                                     \Omega^\#(g_0^2) \times
                                     \ia(g_0^2) ]
\label{eq:cont}
\ee
where $Z^{Ren}$ is the renormalisation constant appropriate for $\Omega$
(which is unity in the case of hadronic masses), and $\Omega^\#$ is the
(dimensionless) lattice value for $\Omega$.
The limit $g_0 \rightarrow 0$ simply expresses the fact that the physical
value is obtained by taking the continuum limit of the lattice estimates.
Assuming that there is no phase transition between values of $g_0^2$
used in present lattice calculations and $g_0=0$, we can write
eq.(\ref{eq:cont}) as

\[
\Omega = [ \lim_{g_0\rightarrow 0}   Z^{Ren}(g_0^2) ] \times
           \lim_{g_0\rightarrow 0} [ \Omega^\#(g_0^2) \times
                                     \ia(g_0^2) ]
\]
For quantities where $Z^{Ren}(g_0^2)$ can be expressed as polynomial in
$g_0^2$ (such as decay constants and hadronic masses) we have

\be
\Omega = Z^{Ren}(g_0^2=0) \times \frac{\Lambda_L}{\lambda_\Omega}
\label{eq:cont2}
\ee
where we have fitted $\ia$ according to 
eq.(\ref{eq:ldpt_fit}) or eq.(\ref{eq:ldpt_fit2}), and we have fitted
$\Omega^\#$ along similar lines as follows,
\[
\frac{1}{\Omega^\#(g_0^2)} = \frac{\lambda_\Omega} {f_{PT}(g_0^2)}
\times  \left[ 1 - \x \frac{g_0^\nu f_{PT}^n(g_0^2)}
                       {           f_{PT}^n(1)    } \right].
\]
Typically $Z^{Ren}(g_0^2=0)$ is known immediately, without the need for any
calculation, (e.g. it is unity for decay constants and hadronic masses).
This means that the continuum estimate of $\Omega$ can be found using
eq.(\ref{eq:cont2}) simply via two lattice distorted PT style fits.

\section*{Acknowledgements}

It is a pleasure to thank
Luigi Del Debbio,
Rajan Gupta,
Simon Hands,
Richard Kenway, 
Vittorio Lubicz,
John Megehan,
Prospero Simonetti,
Stefan Sint,
John Stack
and Tassos Vladikas
for useful discussions.
The author also wishes to acknowledge many useful discussions
with colleagues in the APE and UKQCD collaboration.


\end{document}